\documentclass[fleqn,usenatbib]{mnras}

\usepackage{mathptmx}

\usepackage{amssymb}
\usepackage{xspace}

\usepackage[T1]{fontenc}

\usepackage{graphicx}	% Including figure files
\usepackage{amsmath}	% Advanced maths commands
\usepackage{lineno} % Add line numbers
\usepackage{anyfontsize}
\usepackage{mathtools}

\let\oldequation\equation
\let\oldendequation\endequation

\renewenvironment{equation}
  {\linenomathNonumbers\oldequation}
  {\oldendequation\endlinenomath}

\DeclareMathAlphabet{\mathcal}{OMS}{cmsy}{m}{n}

\newcommand{\txr}{}
\newcommand{\ipr}{\emph{r}\xspace}

\title[Evolution of dark and luminous matter]{Project Dinos II: Redshift evolution of dark and luminous matter density profiles in strong-lensing elliptical galaxies across $0.1 < z < 0.9$}

\author[Sheu et al.]{William Sheu,$^{1}$\thanks{E-mail: wsheu@astro.ucla.edu}
Anowar J.~Shajib,$^{2, 3}$\thanks{NHFP Einstein Fellow}
Tommaso Treu,$^{1}$
Alessandro Sonnenfeld,$^{4}$
Simon Birrer,$^{5}$ 
\newauthor Michele Cappellari,$^{6}$ 
Lindsay J.~Oldham,$^{7}$ 
Chin Yi Tan$^{2, 8}$ 
\\
$^{1}$Department of Physics and Astronomy, University of California, Los Angeles, CA 90095, USA\\
$^{2}$Kavli Institute for Cosmological Physics, University of Chicago, Chicago, IL 60637, USA\\
$^{3}$Department of Astronomy and Astrophysics, University of Chicago, Chicago, IL 60637, USA\\
$^{4}$Department of Astronomy, School of Physics and Astronomy, Shanghai Jiao Tong University, Shanghai 200240, China\\
$^{5}$Department of Physics and Astronomy, Stony Brook University, Stony Brook, NY 1794, USA\\
$^{6}$Subdepartment of Astrophysics, Department of Physics, University of Oxford, Oxford OX1 3RH, UK \\
$^{7}$Institute of Cosmology and Gravitation, University of Portsmouth, Portsmouth PO1 3FX, UK\\
$^{8}$Department of Physics, University of Chicago, Chicago, IL 60637, USA
}

\date{Accepted 2025 June 10. Received 2025 May 28; in original form 2024 August 16}

\pubyear{2025}

\begin{document}
% \linenumbers
\label{firstpage}
\pagerange{\pageref{firstpage}--\pageref{lastpage}}
\maketitle

\begin{abstract}
%The Navarro--Frenk--White profile is well-known to describe unperturbed dark matter halos 
%at high redshifts of $z>3$
%. However, the mass density profile in real galaxies and its dependence on redshift are relatively unknown, as the redistribution of the matter through merger events and other baryonic feedback processes may significantly impact the structure over time. 
We present a new measurement of the dark and luminous matter distribution of massive elliptical galaxies, and their evolution with redshift, by combining strong lensing and dynamical observables. Our sample of \txr{56} lens galaxies covers a redshift range of $0.090 \leq z_{\rm l} \leq 0.884$.  By combining new \textit{Hubble Space Telescope} imaging with previously observed velocity dispersion and line-of-sight measurements, we decompose the luminous matter profile from the dark matter profile and perform a Bayesian hierarchical analysis to constrain the population-level properties of both profiles.  
We find that the inner slope of the dark matter density profile (``cusp''; $\rho_{\rm DM} \propto r^{-\gamma_{\rm in}}$) \txr{is consistent ($\mu_{\gamma_{\rm in}}=0.97^{+0.03}_{-0.03}$ with $\leq0.07$ intrinsic scatter) with a standard Navarro--Frenk--White (NFW; $\gamma_{\rm in}=1$) at $z=0.35$.  Additionally, we find an appreciable evolution with redshift (\txr{$d\log(\gamma_{\rm in})/dz=-0.44^{+0.14}_{-0.15}$}) resulting in a shallower slope (of $> 2 \sigma$ tension from NFW) at redshifts $z \ge 0.49$.  This is in excellent agreement with previous population-level observational studies, as well as with predictions from hydrodynamical simulations such as IllustrisTNG. }
We also find the stellar mass-to-light ratio at the population level is consistent with that of a Salpeter initial mass function, a small stellar mass-to-light gradient ($\kappa_{*}(r)\propto r^{-\eta}$, with \txr{$\overline{\eta} \leq 5 \times 10^{-5}$}), and isotropic stellar orbits. Our averaged total mass density profile is consistent with a power-law profile within 0.25 to 4 Einstein radii (\txr{$\overline{\gamma} = 2.24 \pm 0.14$}), with an internal mass-sheet transformation parameter \txr{$\overline{\lambda} = 0.96 \pm 0.03$} consistent with no mass sheet. Our findings confirm the validity of the standard mass models used for time-delay cosmography. 
%{\bf TT: we need a sentence here to summarize what this means}

%Due to the elusive nature of dark matter, strong lensing remains one of the best methods of quantifying these dark matter halos at higher redshifts. By taking advantage of the numerous advancements in strong lensing detection and modelling in the past few years, it becomes possible to thoroughly explore the evolution of these dark matter halos and redshift.

%we decompose the luminous matter profile from the dark matter profile, and finally constrain dark matter properties and evolution at a population level. 
\end{abstract}


%%%%%%%%%%%%%%%%%%%%%%%%%%%%%%%%%%%%%%%%%%%%%%%%%%

%%%%%%%%%%%%%%%%% BODY OF PAPER %%%%%%%%%%%%%%%%%%

\section{Introduction} \label{sec:introduction}
The progenitors of present-day elliptical galaxies are thought to have formed at $z > 3$, from gas accreting into overdensities in the primordial cosmic web \citep{rees1977, white1991, voort2011}. Dark-matter-only $N$-body simulations predict that matter in these overdensities should distribute itself into a Navarro--Frenk--White (NFW) profile, where the logarithmic density slope scales with $r^{-1}$ in the inner cusp, and $r^{-3}$ outside of the cusp \citep[e.g.,][]{nfw0, nfw1, ghigna2000, diemand2005, gao2012}. However, baryonic processes are thought to actively shape the luminous and dark matter profiles of these elliptical galaxies. Time-accumulated processes, such as adiabatic cooling, mergers, and baryonic feedback, are expected to alter both mass distributions into what we observe in populations of low-$z$ elliptical galaxies \citep[e.g.,][]{silk1998, zant2001, matteo2005, springel2005}.

Dark-matter-only simulations have made great strides in conceptualizing our understanding of these structures.  However, while much can be gleaned through simulations, some inconsistencies still arise when they are compared to observations.  While simulations predict an NFW profile for an unperturbed dark matter distribution, independent analyses of the rotational curves of dwarf galaxies strongly favour a more ``cored'' (i.e., shallower inner slope) density profile \citep{moore1994, flores1994, burkert1995, rhee2004}.  This disagreement, known as the ``core-cusp problem,'' continues to obscure our understanding of dark matter microphysics \citep{blok2010, bullock2017}.  Another problem related to this is how the total mass distributions of massive elliptical galaxies seem consistent with a power-law profile $\rho = r^{-\gamma}$ within their effective (i.e., half-light) radius, which would imply that the luminous and dark matter profiles combine to form a power-law despite neither innately conforming to this profile. This phenomenon is known as the ``bulge-halo conspiracy'' \citep{D+T14}.  This nearly isothermal trend, with small scatter $\mu_{\gamma} = 2.19 \pm 0.03$ and $\sigma_\gamma = 0.11$ \citep{Cappellari2015}, was found to extend out to about four times the Einstein radius in elliptical galaxies with large velocity dispersion \citep{Serra2016, sahu2024}. However, when studies were extended to larger samples at lower velocity dispersion $\sigma_{\rm v}$, the ``universal'' total slope was found to be part of a trend, which flattens at around $\log\sigma_{\rm v} \approx 2.1$ ($\sigma_{\rm v}$ in units of km s$^{-1}$) in elliptical galaxies \citep{Poci2017}, reaching $\langle \gamma \rangle = 1.5$ around $\log\sigma_{\rm v} \approx 1.8$ for samples including spiral galaxies \citep{Li2019} with a clear age dependence, resulting in lower $\gamma$ for younger galaxies at fixed $\sigma_{\rm v}$ \citep{Zhu2024}. Keeping to the cold dark matter regime, a possible explanation of these observations is that the baryonic processes gradually influence the inner dark matter density slope.  These results also suggest that the galaxies’ detailed formation history affects this process.  Therefore, it is helpful to quantify how the dark matter profile evolves with redshift; in other words, how time-accumulated baryonic processes could potentially alter its shape.

% We adopt a generalized NFW profile \citep[gNFW;][]{keeton2001, wyithe2001} in this paper, to account for how dark matter halos evolve over redshift through baryonic processes.  The gNFW density profile is defined as: 
% %
% \begin{equation}\label{gnfw_eq}
% \rho (r) = \frac{\rho_{\rm s}}{(r/R_{\rm s})^{\gamma_{\rm in}}(1+r/R_{\rm s})^{3-\gamma_{\rm in}}},
% \end{equation}
% % \rho (r) = \Sigma_{\text{crit}} \frac{\kappa_{\rm s} / R_{\rm s}}{(r/R_{\rm s})^{\gamma_{\rm in}}(1+r/R_{\rm s})^{3-\gamma_{\rm in}}},
% %where $\Sigma_{\text{crit}}$ is the lensing critical surface mass density, $\kappa_{\rm s}$ is the convergence at $R_{\rm s}$,
% where $R_{\rm s}$ is the scale radius, $\rho_{\rm s}$ is the density at $R_{\rm s}$, and $\gamma_{\rm in}$ is the logarithmgic inner slope.  With $\gamma_{\rm in} = 1$, the gNFW profile becomes an NFW profile.  This parameterization of $\gamma_{\rm in}$ can serve as a proxy for dark matter halo evolutionary processes (mergers, stellar/supernovae feedback, etc), with $\gamma_{\rm in} < 1$ roughly describing an adiabatically-expanded cored halo and $\gamma_{\rm in} > 1$ roughly describing an adiabatically-contracted cuspy halo \citep[e.g.,][]{gnedin2011, newman2013}.  %In this paper, we measure the redshift evolution of $\gamma_{\rm in}$, which describes how "cored" ($\gamma_{\rm in} < 1$) or "cuspy" ($\gamma_{\rm in} > 1$) a dark matter halo is. 

Understanding the distribution of dark matter within elliptical galaxies is further complicated by the need to separate it from the baryonic component, which is dominated by stars in their centre. Thus, the observational characterization of dark matter halos is intertwined with the stellar mass-to-light properties of these galaxies.  This strongly depends on the presence of low-mass stars and high-mass stellar remnants, whose presence can only be inferred indirectly via galaxy stellar dynamics \citep[e.g.,][]{Cappellari2012,Cappellari2013,O+A18,Mehrgan24,Lu2024}, a combination of dynamics and lensing probes \citep[e.g.,][]{Treu10,Auger2010imf}, microlensing \citep[e.g.,][]{Schechter14,JVM19}, or faint stellar absorption features in the galaxy spectra \citep[e.g.,][]{Dokkum2010,Spiniello2012,Labarbera17,Lagattuta17}. See reviews by \citet{Smith20} and \citet{Cappellari16}.  The stellar mass-to-light ratio and gradient (formulations shown in Section~\ref{sec:hierarchial}) are two parameters that have been shown to accurately trace the presence of stellar mass within elliptical galaxies.

As dark matter structures cannot be directly observed, strong lensing systems are invaluable tools for probing invisible structures at redshift $z \gtrsim$ 0.1.  By measuring and modelling the imaging configuration of strongly lensed systems, it is possible to constrain the total mass profile of the lens galaxy \citep[see reviews by][]{treu2010, shajib2024}.  Folding in additional kinematic and line-of-sight (LOS) measurements then allows for an accurate decomposition of the dark and luminous matter profiles in the lens galaxies \citep[e.g.,][]{treu2004,newman2013, shajib2021}.

With a better understanding of the matter distributions and evolution in galaxies, tighter constraints on cosmology through time delay cosmography \citep{treu2016} can also be achieved.  The flat $\Lambda$ Cold Dark Matter ($\Lambda$CDM) cosmological model is highly successful in explaining observables from the time of photon decoupling (at $z \approx$ 1100) to the present time \citep{planck2020}.  According to this model, our Universe has a flat geometry and is expanding at an accelerating rate \citep{riess1998, perlmutter1999}.  The inferred value for $H_0$ (the present-day expansion rate of the Universe, or the Hubble constant) from the \textit{Planck} CMB measurements is $67.4\pm 0.5$~km s$^{-1}$ Mpc$^{-1}$ \citep{planck2020}.  In contrast, direct measurements of $H_0$ using cosmic distance ladders of Type Ia supernovae calibrated with Cepheids are higher by $\gtrsim 5\sigma$ \citep{riess2022,Abdalla22}. Thus, if this inconsistency is not due to systematic effects, then the $\Lambda$CDM model would require revisions \citep{valentino2021}. 

Competitive late-time $H_0$ constraints from time-delay cosmography can be attained if the time delays can be measured precisely and the lensing potential can be modelled accurately, providing an independent method of measuring the Hubble constant \citep[e.g.,][]{treu2022,treu2023, birrer2024}.  This is most commonly achieved by identifying and analyzing strongly lensed quasar systems \citep[e.g.,][]{lemon2018, lemon2022, sheu2024}, or more rarely using strongly lensed supernovae \citep[e.g.,][]{quimby2014, sheu2023, pascale2025}.  The mass-sheet degeneracy \citep{Falco1985,S+S13}, however, is currently the main source of residual uncertainty in these measurements \citep{birrer2020}.  Therefore, information about the lens mass profile, such as that obtained in this paper, is crucial to mitigate the mass-sheet degeneracy and tighten the constraints on $H_0$ \citep{birrer2020b}. 

% This is the second paper of the Project Dinos series (PI: Shajib)\footnote{\textit{HST} archival program AR-16149}, which aims to study the evolution of dark and luminous matter distributions within elliptical galaxy lenses.  In the first paper, \citet[][henceforth referred to as Dinos-I]{dinos1} uniformly model a sample of 77 galaxy-scale lens, in order to investigate the validity of using a power-law profile to represent the total-matter surface density profile in elliptical galaxies.

In this Project Dinos\footnote{\url{https://www.projectdinos.com}} paper, we investigate the radial density profiles of the dark matter and baryons for a sample of elliptical lens galaxies at redshifts $0.1 < z < 0.9$. Our goal is to improve our understanding of the properties and evolution of massive elliptical galaxies.
%, with emphasis on the growth of their dark matter halos and the initial mass function of their stellar populations.
Specifically, we aim to constrain how the inner slope of the dark matter profile evolves with redshift, investigate correlations between dark and luminous matter density profiles with velocity dispersion, determine the normalization and gradient of the stellar mass-to-light ratio to study the stellar initial mass function, and test the consistency of the total matter surface density with a power-law profile. 

Our full sample of 56 lens galaxies consists of 33 lenses from the Sloan Lens ACS \citep[SLACS;][]{slacs} survey, and 23 lenses from the Strong Lensing Legacy Survey \citep[SL2S;][]{sl2s,gavazzi12}. Among these, we present power-law lens models of 21 SL2S systems in this paper from newly obtained high-resolution imaging from the \textit{Hubble Space Telescope} (\textit{HST}). These new observations have a higher signal-to-noise ratio ($S/N$) compared to previous \textit{HST} observations available for 12 out of the 21 systems, and the remaining nine of the 21 have been observed with the \textit{HST} for the first time. This deep and high-$S/N$ imaging data allows for tighter constraints on their mass distribution properties. For the remaining systems in our full sample (33 SLACS and two SL2S lenses), we adopt the power-law lens models based on archival \textit{HST} imaging from \citet[][henceforth referred to as Dinos-I]{dinos1}. Dinos-I performed a joint lensing--dynamics analysis based on these lens models to find that the power-law profile can describe the total mass distribution within $1\sigma$ statistical consistency. Although Dinos-I attempted to constrain any evolutionary in the deviation (or, the lack thereof) from the power-law description, the lower $S/N$ of the archival \textit{HST} imaging for the high-redshift ($z\sim0.6$) SL2S sample prevented a tight constraint on the evolutionary parameters. In this paper, we resolve this issue by obtaining and modelling deeper imaging of the SL2S sample.  The knowledge acquired about the mass density profile of elliptical galaxies and the stellar anisotropy will inform $H_0$ measurements via time-delay cosmography. To that end, we perform a Bayesian hierarchical analysis jointly on lensing and dynamical observables to constrain population-level properties pertaining to dark matter and baryonic distributions in our lens galaxy sample.

\begin{figure*} 
    \includegraphics[width=1\linewidth]{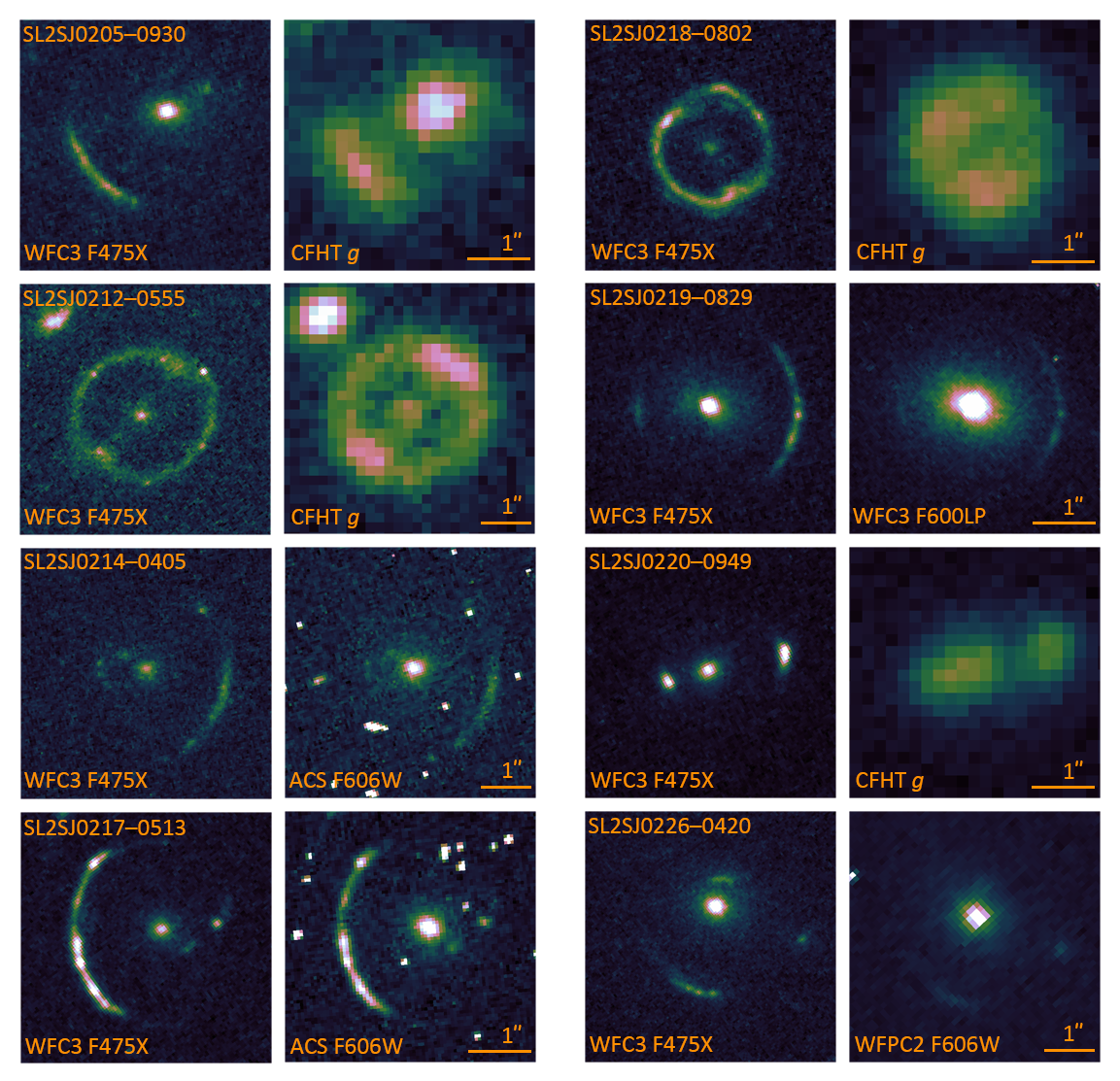}
    \caption{Comparison of the first eight (of 21) SL2S systems observed in \textit{HST}-GO-17130, to prior \textit{HST} or CFHT imaging.  For each pair of images, the WFC3 F475X image is shown on the left, and prior \textit{HST} imaging (if available; otherwise, CFHT~$g$-band imaging) is shown on the right.  All images are orientated such that the North is up and the East is left.  Qualitatively, the F475X images are many times sharper than previous imaging for a large majority of the systems.}
    \label{fig:sl2s_comparision0}
\end{figure*}

\begin{figure*}
    \includegraphics[width=1\linewidth]{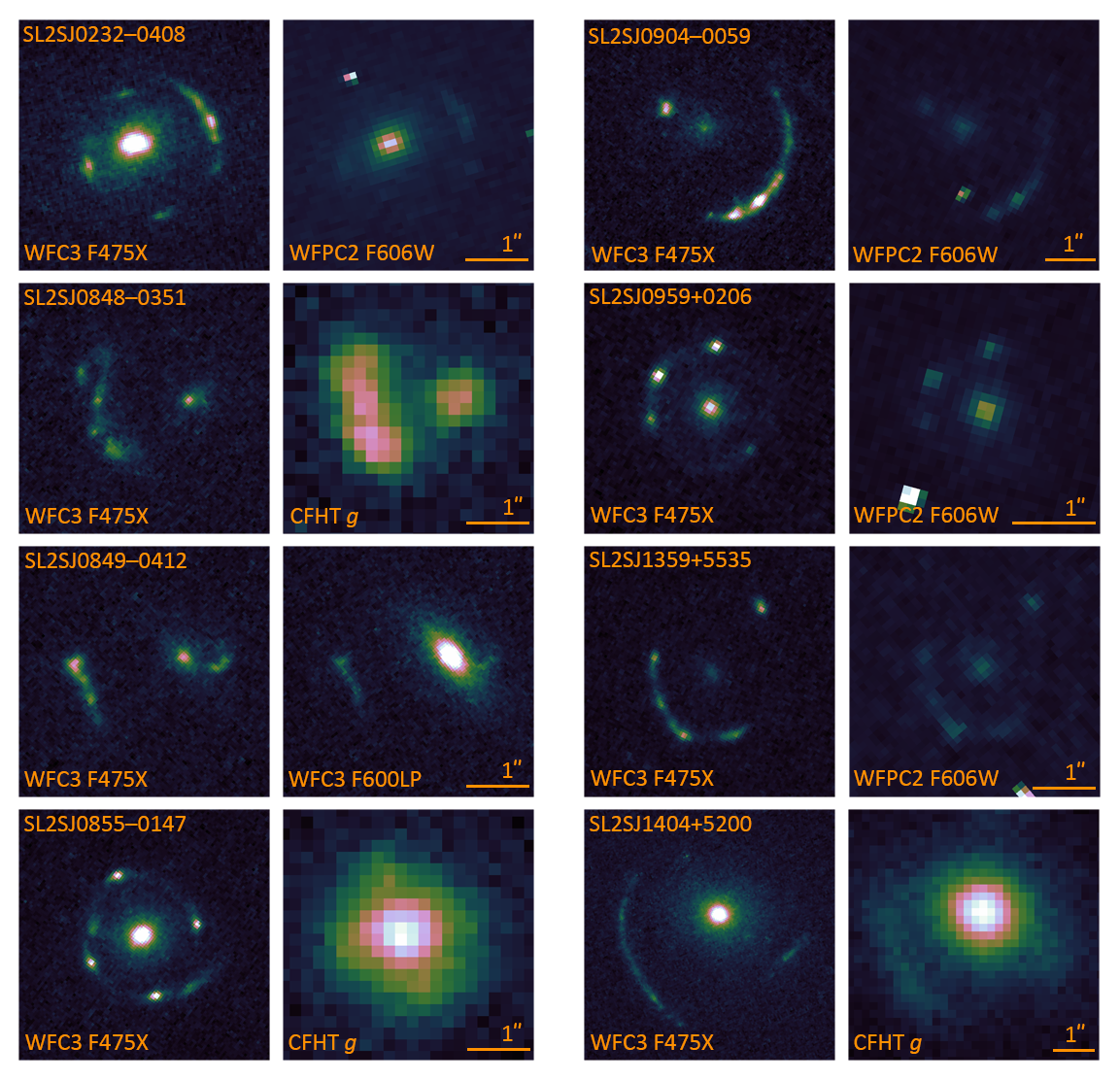}
    \caption{Comparison of the next eight (of 21) SL2S systems observed in \textit{HST}-GO-17130, to prior \textit{HST} or CFHT imaging.  See the caption of Figure~\ref{fig:sl2s_comparision0} for a full description.}
    \label{fig:sl2s_comparision1}
\end{figure*}

\begin{figure*}
    \includegraphics[width=1\linewidth]{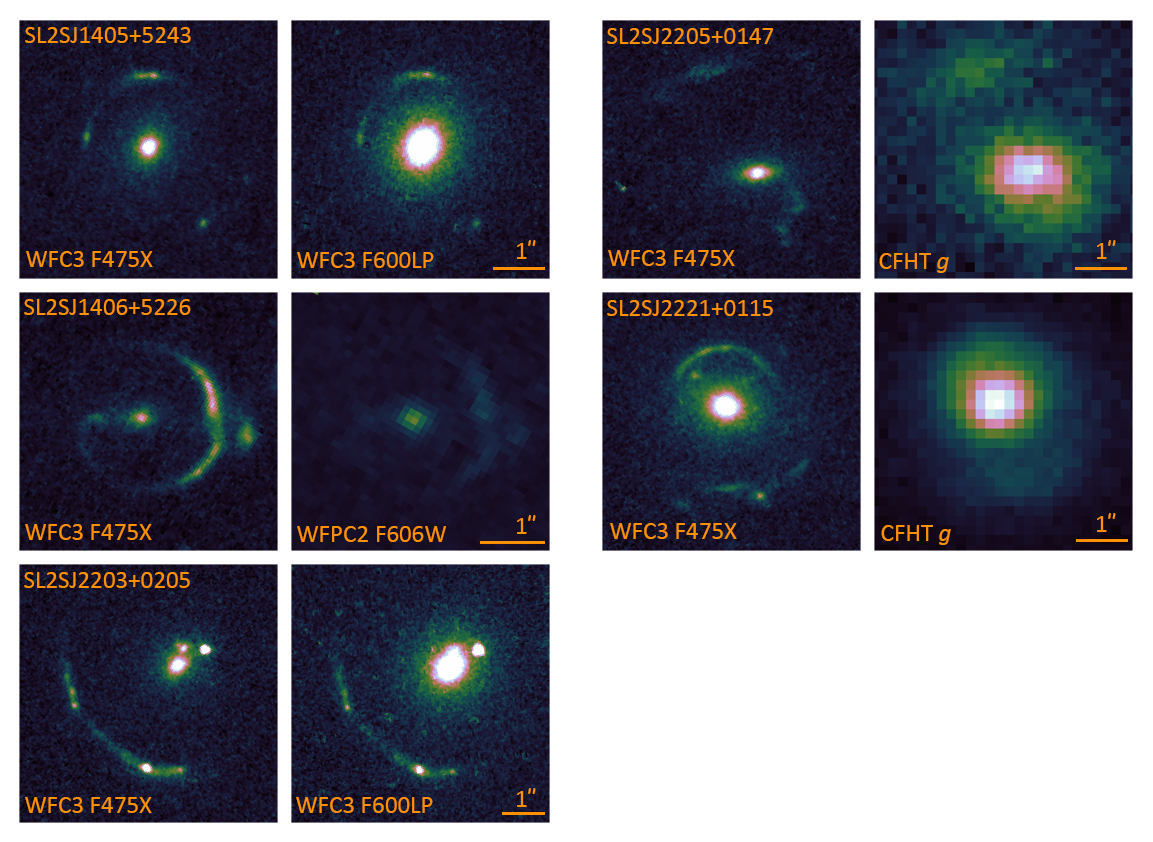}
    \caption{Comparison of the last five (of 21) SL2S systems observed in \textit{HST}-GO-17130, to prior \textit{HST} or CFHT imaging.  See the caption of Figure~\ref{fig:sl2s_comparision0} for a full description.}
    \label{fig:sl2s_comparision2}
\end{figure*}

We organize this paper as follows. In Section~\ref{sec:observation}, we introduce our full lens sample.  We then outline the modelling process for the new SL2S system in Section~\ref{sec:modelling}.  Next, we describe our hierarchical Bayesian inference pipeline to constrain population level statistics of our full sample in Section~\ref{sec:hierarchial}. We present the results of our analysis in Section~\ref{sec:results}, and discuss their implications for galaxy evolution, our understanding of elliptical galaxies, and cosmography in Section~\ref{sec:discussion}.  Finally, we conclude our findings in Section~\ref{sec:conclusion}.  Throughout the paper, we adopt a flat $\Lambda$CDM cosmology model \txr{based on the Wilkinson Microwave Anisotropy Probe (WMAP) 5-year observations \citep{Komatsu2009}, as our assumption of the mass-concentration relation is derived from these values \citep{duffy2008}.  This includes $w_0 = -1$, $\sigma_8 = 0.796$, $\Omega_{\rm m}=0.3$, $\Omega_{\rm b}=0.047$, and $H_0 =70$ km s$^{-1}$ Mpc$^{-1}$, unless stated otherwise.
% with $\Omega_{\rm m}=0.3$, $\Omega_{\rm b}=0.047$, $H_0 =70$ km s$^{-1}$ Mpc$^{-1}$, unless stated otherwise.  \txr{We also adopt $\sigma_8 = 0.796$ and $w_0 = -1$ from the Wilkinson Microwave Anisotropy Probe (WMAP) 5-year observations \citep{Komatsu2009}, as our assumption of the mass-concentration relation is derived from these values \citep{duffy2008}.  
Our choice in cosmological parameters is motivated by how the mass-concentration relation is most significantly affected by the $\sigma_8$ and $w_0$ parameters, while being fairly robust against $\Omega_{\rm m}$, $\Omega_{\rm b}$, and $H_0$ \citep[e.g.,][]{lopez-cano2022}.}% Our findings do not depend significantly on the choice of cosmological parameters. 
% Any $\log()$ in this paper should be assumed to be of base 10.

\section{Lens sample and data} \label{sec:observation}
% (https://www.stsci.edu/hst/phase2-public/17130.pdf)
% \begin{equation} 
%     \frac{r_f}{r_i} = \frac{M_f(r_f) - M_{{\rm gal}, i}(r_f)}{(1-f_{\rm gal})M_f(r_f)}
% \end{equation}

% \begin{equation} 
%     f_{{\rm gal}} = \frac{M_{{\rm gal}, i}}{M_{{\rm total}, i}}
% \end{equation}
Our full sample in this paper consists of 56 elliptical galaxy lenses.  Of these, 33 are from the SLACS sample and 23 from the SL2S sample, spanning a lens redshift range of $0.090 \leq z_{\rm l} \leq 0.884$, with a mean redshift of 0.353 and a median redshift of 0.277.  We use the lens models and lens galaxies’ light profiles of all 33 SLACS and two SL2S systems from Dinos-I \txr{(excluding systems without reliable lens and source redshifts)}, modelled using archival \textit{HST} imaging.  

% \txr{There are also 33 SLACS and four SL2S lens models and lens galaxies' light profiles from Dinos-I, modelled using archival \textit{HST} imaging.  We utilize these additional models into our hierarchical analysis, with the exception of two of the four SL2S systems due to large uncertainties in their velocity dispersions (see below).}

% In addition to these 21 SL2S systems, we also include four additional SL2S and 33 SLACS strong lenses utilizing the lens surface density and light profiles previously modelled in Dinos-I.  This extends our full sample's redshift range to $z_{\rm l}=[0.090, 0.884]$, with $\langle z \rangle= 0.347$ and a total of 58 systems.   

We model the remaining 21 SL2S lenses from our full sample, using newly observed, high-resolution \textit{HST} imaging in the F475X filer obtained through the program HST-GO-17130 (PI: Treu), nine of which were previously unobserved by \textit{HST}. Although archival \textit{HST} images exist for many of these systems, they have significantly shorter exposure times, making them comparatively less powerful for constraining lens models and studying lens galaxy structures (Dinos-I). The new images presented in the paper were obtained using the Wide Field Camera 3 (WFC3), with at least 2366 seconds of exposure per system from 2023 January 9 to 2024 April 14. The wide and blue F475X filter was selected to observe the lensed blue arcs at a high $S/N$, while minimizing contamination by the redder lens galaxy's light. A comparison between the F475X observations and previously observed \textit{HST} or Canada--France--Hawaii Telescope (CFHT) imaging is presented in Figures~\ref{fig:sl2s_comparision0}, \ref{fig:sl2s_comparision1}, and \ref{fig:sl2s_comparision2}. These 21 SL2S systems uniformly sample lens redshifts $0.3 < z_{\rm l} < 0.9$.
These 21 systems were selected out of the full SL2S sample of 35 lenses because of the availability of redshifts for both the lensing and source galaxies and the velocity dispersions of the lens galaxies \citep{sonnenfeld2013, sonnenfeld2015}.  In this paper, we present lens models using these new and deeper observations for the 21 SL2S lenses, which are of higher quality than those previously published (\citealp{sonnenfeld2013, sonnenfeld2015}; Dinos-I).

% \txr{We incorporate the LOS measurements of the SL2S systems \citep{wells2024} and the SLACS systems \citet{birrer2020} into the hierarchical analysis (see Section \ref{sec:hierarchial}).  For t}
\txr{We incorporate the velocity dispersion \citep{Mozumdar2025} and the LOS measurements \citep{wells2024} of the SL2S systems into the hierarchical analysis (see Section \ref{sec:hierarchial}).  %From Mozumdar et al. 2025 (in preparation), we found that two of the four SL2S systems from Dinos-I sample (which do not have new F475X \textit{HST} imaging) had their velocity dispersions measured only by the Low Resolution Imaging Spectrometer \citep[LRIS;][]{Oke1995} blue channel, which has a very low resolution (resulting in uncertainties of $\sim 200$ km/s).  As such, we exclude these two systems (SL2SJ0225-0454 and SL2SJ1411+5651) from our hierarchical analysis. 
The velocity dispersions for the SLACS systems are provided by \citet{knabel2025}, and the LOS information by \citet{birrer2020}.  All velocity dispersion measurements have been thoroughly analyzed, and have both their systematics and statistical uncertainties assessed \citep{knabel2025,Mozumdar2025}.}

While the F475X filter provides high $S/N$ on the lensed arcs and hence tight constraints on the lensing parameters, on average it contains less of the lens galaxy's light compared to redder filters.  Also, this filter is much bluer compared to the bands used to measure the velocity dispersions.  Therefore, for the lens light profile used in the dynamical modelling, we use the lens galaxy's light profile derived from the \ipr-band CFHT observations after tightly constraining the flux distribution on the lensed arcs using our lens models and source reconstructions based on the \textit{HST} imaging (see Section~\ref{cfht}).  

% In addition to these 21 SL2S systems, we also include four additional SL2S and 33 SLACS strong lenses utilizing the lens surface density and light profiles previously modelled in Dinos-I.  This extends our full sample's redshift range to $z_{\rm l}=[0.090, 0.884]$, with $\langle z \rangle= 0.347$ and a total of 58 systems. The velocity dispersions for the SLACS systems are provided by \citet{shu2015}, and LOS information by \citet{birrer2020}.  

\section{Lens galaxy's mass and light distributions} \label{sec:modelling}
In this section, we first present our general lens modelling pipeline, applied to the new \textit{HST} imaging for 21 SL2S systems in Section~\ref{sec:lens_modelling}. We then explain the modelling process for the CFHT \ipr-band light profiles in Section~\ref{cfht}. Lastly, we describe how we account for systematic uncertainties in Section~\ref{sec:systematic}.

\subsection{Lens modelling} \label{sec:lens_modelling}

We use \textsc{lenstronomy}\footnote{\url{https://github.com/lenstronomy/lenstronomy}}, a multi-purpose lens modelling software package, to model our lenses from the F475X data \citep{birrer2018, lenstronomyII}.  \textsc{lenstronomy} has proven to be reliable and robust in the Time-Delay
Lens Modelling Challenge \citep[TDLMC;][]{ding2021}, where \textsc{lenstronomy} was used by two independent teams to recover lens model parameters within statistical consistency (in Rung 2).  

We base our modelling pipeline on previous efforts by \citet{Shajib19} and \citet{Schmidt23}. A few of our systems require additional modifications to the pipeline (e.g., including an additional source light profile, lens light profile, and/or lensing profile). See Appendix~\ref{appendix:models} for system-specific amendments to the general modelling pipeline.  Our procedure for lens modelling consists of using a particle swarm optimization operation \citep{Kennedy1995} to locate a maximum of the lens likelihood function likely close to the global, and then a Markov chain Monte Carlo (MCMC) algorithm with \textsc{emcee} to obtain the posterior of the model parameters \citep{emcee}.  %For modelling the newly obtained \textit{HST} data, we use \textsc{lenstronomy} to constrain the strong-lensing parameters.  %For modelling the CFHT \ipr-band observations, we use \textsc{lenstronomy} to constrain the lens light parameter, with tight lensing constraints from the F475X model.

We model the lens light using the elliptical de Vaucouleurs profile \citep{vaucouleurs1948}, which is the special case of the S\'ersic profile with the S\'ersic index fixed at $n_{\rm s} = 4$ \citep{sersic}. We opt to use a simple lens light model, fixing $n_{\rm s} = 4$, as the lens galaxy light is not very extended relative to the arc position in the observed F475X band. Hence, contamination of the lens light to the source light becomes much less an issue, and a simpler lens light model would prevent the lens light from overfitting for the oftentimes brighter source galaxy light.  %Additionally, since we ultimately use the CFHT \ipr-band for the lens light profile in the subsequent dynamical modelling (Section~\ref{cfht}), we do not find it necessary to utilize an overly flexible lens light profile.
%Firstly, we model the lens light using one or two elliptical Sérsic light profile(s), selected based on a lower Bayesian information criterion (BIC).  In the case of 1 Sérsic profile, the Sérsic index has an uniform prior from \txr{$n_{\rm s} = [\_, \_]$}.  In the case of two Sérsic profiles, the Sérsic indices are fixed to $n_{\rm s} = 1$ and $n_{\rm s} = 4$, to simulate an exponential profile and de Vaucouleurs profile respectively \citep{vaucouleurs1948}.  The centres and ellipticities are the same in the case of a two Sérsic profile.

The lens model utilizes a power-law elliptical mass distribution \citep[PEMD,][]{Barkana1998}, and an external shear component.  We impose a Gaussian prior on the centre of the lens profile, with the mean set at the lens light model centre and a standard deviation of $0.04''$, or one pixel in the F475X image.  We also impose prior conditions on the lens profile ellipticity as $q_{\rm m} \ge q_{\rm L} - 0.1$ and \txr{$|{\Delta}_{\rm PA}| \le 10-5/(q_{\rm L} -1)$, where $q_{\rm m}$ is the mass axis ratio, $q_{\rm L}$ is the light axis ratio, and ${\Delta}_{\rm PA}$ the difference between the position angles of the mass and light profile \citep{Schmidt23}.  These well justified priors are to prevent nonphysical configurations such as an extraordinarily large ellipticity in the mass distribution, and rather shifts its effects onto the external shear component.}  The source light model utilizes a basis set that includes an elliptical S\'ersic profile and shapelets \citep{shapelets1, shapelets2}. The number of shapelets is determined by the configuration with the lowest Bayesian information criterion (BIC). %by incrementally increasing the shapelet order parameter $n_{\rm max}$ from zero. When the Bayesian information criterion (BIC) begins to increase relative to the the previous order, we select the order with the lowest BIC.
We ensure the convergence of the MCMC chains according to the method as follows. After running the chain for an initial 200 steps, we start monitoring the median and spread (16th and 84th percentiles) for each model parameter.  When all the model parameter means at a given step do not change from that of the previous 100 steps by more than five per cent of the spread, we consider the chains to have converged \citep{Schmidt23}.

\begin{figure*}
    \includegraphics[width=0.95\linewidth]{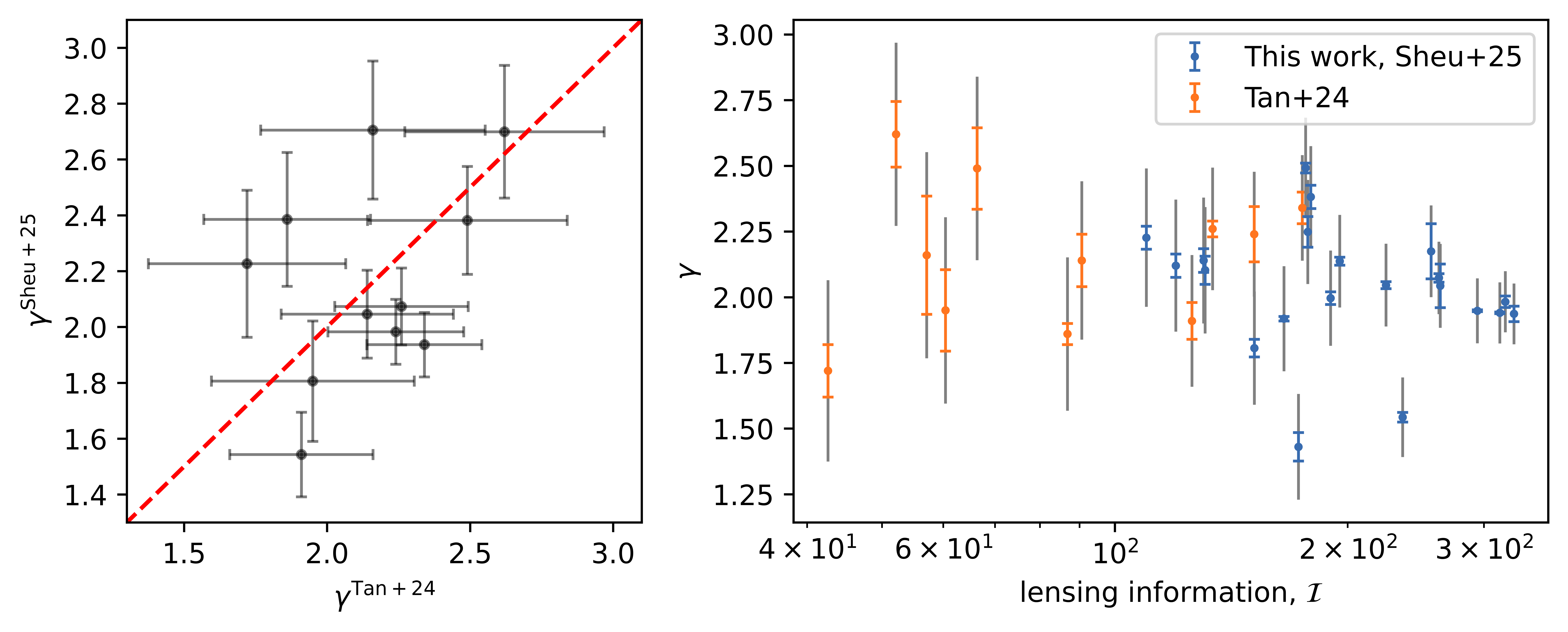}
    \caption{Comparison of logarithmic slopes ($\gamma$) between our and Dinos-I's models (Sheu+25 and Tan+24, respectively).  On the left plot, we compare Dinos-I and our values of $\gamma$ (on the x and y axes, respectively), for the 11 overlapping systems between our samples.  The grey error bars represent the systematic uncertainties, and the one-to-one relation is traced with the red dashed line.  We find that our measurements generally match that of Dinos-I, with tighter uncertainties.  On the right plot, we plot lensing information $\mathcal{I}$ (defined in Equation \ref{eq:lensing_info}) versus $\gamma$ for our 21 systems (in blue) and the 11 overlapping systems from Dinos-I (in orange). The data for the overlapping systems in Dinos-I has less lensing information due to their shorter exposure times from a snapshot program, and their comparatively redder bands used. The coloured uncertainties are the statistical uncertainties, whereas the grey uncertainties are the total ones.  Overall, our sample of SL2S lenses contains more lensing information per system than that in Dinos-I.}
    \label{fig:gamma_systematics}
\end{figure*}

\begin{table*} 
 \caption{SL2S modelling results.  Here, $z_{\rm l}$ is the spectroscopic lens galaxy redshift, $R_\text{E}$ is the Einstein radius, $\gamma$ is the logarithmic slope of the mass profile, $q_\text{m}$ is the mass axis ratio, $\text{PA}_\text{m}$ is the mass position angle, $\gamma^{\text{shear}}$ is the residual shear magnitude, $\phi^{\text{shear}}$ is the residual shear angle, $R_{\text{eff}}$ is the effective or half-light radius in the CHFT \ipr-band, $q_\text{L}$ is the light axis ratio, and $\text{PA}_\text{L}$ is the light position angle.  All degree measures are given as North of East.}
 \label{tab:results}
 \begin{tabular}{ccccccccccc}
  \hline
  Name &    $z_{\rm l}$ &                      $R_\text{E}$ & $\gamma$ &                   $q_\text{m}$ &            $\text{PA}_\text{m}$ &      $\gamma^{\text{shear}}$ & $\phi^{\text{shear}}$ &               $R_{\text{eff}}$ &                      $q_\text{L}$ &                   $\text{PA}_\text{L}$ \\
   & & {[}arcsec{]} & & & {[}\degr{]} & & {[}\degr{]} & {[}arcsec{]} & & {[}\degr{]} \\
  \hline
SL2SJ0205$-$0930 &  0.557 &  $1.246^{+0.009}_{-0.008}$ &  $2.00^{+0.09}_{-0.08}$ &  $0.63^{+0.04}_{-0.03}$ &   $-26^{+2}_{-2}$ &  $0.06^{+0.01}_{-0.01}$ &    $15^{+7}_{-9}$ &  $0.35^{+0.02}_{-0.03}$ &  $0.63^{+0.04}_{-0.04}$ &    $-16^{+4}_{-4}$ \\ SL2SJ0212$-$0555 &  0.750 &  $1.238^{+0.004}_{-0.004}$ &  $1.43^{+0.13}_{-0.11}$ &  $0.70^{+0.01}_{-0.01}$ &   $-26^{+1}_{-1}$ &  $0.14^{+0.01}_{-0.01}$ &   $-20^{+1}_{-1}$ &  $0.29^{+0.06}_{-0.05}$ &  $0.54^{+0.14}_{-0.12}$ &   $-35^{+10}_{-6}$ \\ SL2SJ0214$-$0405 &  0.609 &  $1.340^{+0.008}_{-0.008}$ &  $2.39^{+0.21}_{-0.18}$ &  $0.75^{+0.02}_{-0.03}$ &    $47^{+5}_{-6}$ &  $0.17^{+0.01}_{-0.02}$ &    $82^{+1}_{-1}$ &  $0.62^{+0.06}_{-0.05}$ &  $0.81^{+0.05}_{-0.06}$ &     $28^{+8}_{-7}$ \\ SL2SJ0217$-$0513 &  0.646 &  $1.272^{+0.014}_{-0.014}$ &  $1.94^{+0.12}_{-0.08}$ &  $0.85^{+0.01}_{-0.01}$ &   $-21^{+1}_{-1}$ &  $0.10^{+0.01}_{-0.01}$ &   $-87^{+1}_{-1}$ &  $0.38^{+0.03}_{-0.03}$ &  $0.66^{+0.07}_{-0.08}$ &    $-47^{+7}_{-9}$ \\ SL2SJ0218$-$0802 &  0.884 &  $0.850^{+0.016}_{-0.017}$ &  $2.04^{+0.16}_{-0.14}$ &  $0.88^{+0.01}_{-0.01}$ &    $45^{+3}_{-3}$ &  $0.02^{+0.01}_{-0.01}$ &   $-32^{+6}_{-5}$ &  $0.41^{+0.08}_{-0.08}$ &  $0.51^{+0.22}_{-0.13}$ &    $63^{+7}_{-12}$ \\ SL2SJ0219$-$0829 &  0.389 &  $1.296^{+0.038}_{-0.038}$ &  $2.70^{+0.08}_{-0.13}$ &  $0.67^{+0.03}_{-0.04}$ &    $11^{+2}_{-3}$ &  $0.05^{+0.01}_{-0.01}$ &    $64^{+2}_{-3}$ &  $0.47^{+0.03}_{-0.02}$ &  $0.64^{+0.01}_{-0.02}$ &      $9^{+2}_{-2}$ \\ SL2SJ0220$-$0949 &  0.572 &  $0.914^{+0.023}_{-0.020}$ &  $2.17^{+0.16}_{-0.19}$ &  $0.77^{+0.07}_{-0.08}$ &  $-24^{+7}_{-12}$ &  $0.07^{+0.04}_{-0.03}$ &  $-19^{+7}_{-14}$ &  $0.20^{+0.01}_{-0.01}$ &  $0.62^{+0.05}_{-0.07}$ &    $-29^{+6}_{-3}$ \\ SL2SJ0226$-$0420 &  0.494 &  $1.152^{+0.009}_{-0.009}$ &  $2.05^{+0.14}_{-0.15}$ &  $0.72^{+0.03}_{-0.02}$ &    $21^{+3}_{-3}$ &  $0.07^{+0.01}_{-0.01}$ &    $22^{+4}_{-4}$ &  $0.52^{+0.04}_{-0.04}$ &  $0.78^{+0.03}_{-0.04}$ &     $34^{+6}_{-5}$ \\ SL2SJ0232$-$0408 &  0.352 &  $1.019^{+0.019}_{-0.019}$ &  $2.38^{+0.14}_{-0.23}$ &  $0.70^{+0.01}_{-0.01}$ &   $-23^{+2}_{-2}$ &  $0.05^{+0.01}_{-0.01}$ &   $-18^{+3}_{-3}$ &  $0.53^{+0.03}_{-0.02}$ &  $0.65^{+0.02}_{-0.02}$ &    $-23^{+2}_{-2}$ \\ SL2SJ0848$-$0351 &  0.682 &  $0.930^{+0.038}_{-0.038}$ &  $1.95^{+0.08}_{-0.08}$ &  $0.43^{+0.01}_{-0.01}$ &   $-25^{+1}_{-1}$ &  $0.01^{+0.01}_{-0.01}$ &  $35^{+12}_{-11}$ &  $0.39^{+0.04}_{-0.04}$ &  $0.32^{+0.03}_{-0.04}$ &    $-30^{+4}_{-3}$ \\ SL2SJ0849$-$0412 &  0.722 &  $1.134^{+0.014}_{-0.011}$ &  $2.07^{+0.09}_{-0.22}$ &  $0.71^{+0.03}_{-0.07}$ &   $40^{+10}_{-7}$ &  $0.03^{+0.03}_{-0.02}$ &  $65^{+11}_{-43}$ &  $0.27^{+0.02}_{-0.02}$ &  $0.42^{+0.06}_{-0.05}$ &     $49^{+4}_{-4}$ \\ SL2SJ0855$-$0147 &  0.365 &  $0.956^{+0.034}_{-0.034}$ &  $2.25^{+0.12}_{-0.10}$ &  $0.83^{+0.01}_{-0.01}$ &   $-22^{+1}_{-1}$ &  $0.04^{+0.01}_{-0.01}$ &    $75^{+1}_{-1}$ &  $0.32^{+0.03}_{-0.02}$ &  $0.80^{+0.02}_{-0.03}$ &    $-18^{+9}_{-7}$ \\ SL2SJ0904$-$0059 &  0.611 &  $1.411^{+0.005}_{-0.005}$ &  $1.98^{+0.14}_{-0.16}$ &  $0.96^{+0.01}_{-0.01}$ &   $12^{+8}_{-12}$ &  $0.08^{+0.01}_{-0.01}$ &   $-59^{+1}_{-1}$ &  $0.30^{+0.02}_{-0.02}$ &  $0.90^{+0.05}_{-0.09}$ &  $-33^{+7}_{-15}$ \\ SL2SJ0959$+$0206 &  0.552 &  $0.711^{+0.010}_{-0.010}$ &  $2.12^{+0.18}_{-0.18}$ &  $0.69^{+0.01}_{-0.01}$ &    $71^{+1}_{-1}$ &  $0.09^{+0.01}_{-0.01}$ &   $83^{+1}_{-1}$ &  $0.41^{+0.05}_{-0.04}$ &  $0.84^{+0.06}_{-0.07}$ &   $62^{+16}_{-23}$ \\ SL2SJ1359$+$5535 &  0.783 &  $1.092^{+0.006}_{-0.006}$ &  $1.81^{+0.07}_{-0.10}$ &  $0.73^{+0.02}_{-0.02}$ &    $62^{+2}_{-2}$ &  $0.09^{+0.01}_{-0.01}$ &  $86^{+3}_{-5}$ &  $0.60^{+0.08}_{-0.08}$ &  $0.50^{+0.10}_{-0.10}$ &     $58^{+7}_{-7}$ \\ SL2SJ1404$+$5200 &  0.456 &  $2.394^{+0.011}_{-0.011}$ &  $2.49^{+0.18}_{-0.24}$ &  $0.75^{+0.01}_{-0.01}$ &    $24^{+1}_{-1}$ &  $0.17^{+0.01}_{-0.01}$ &   $-74^{+1}_{-1}$ &  $0.77^{+0.03}_{-0.02}$ &  $0.79^{+0.02}_{-0.02}$ &     $22^{+3}_{-3}$ \\ SL2SJ1405$+$5243 &  0.526 &  $1.457^{+0.007}_{-0.007}$ &  $2.23^{+0.21}_{-0.14}$ &  $0.69^{+0.01}_{-0.01}$ &   $-41^{+2}_{-2}$ &  $0.07^{+0.01}_{-0.01}$ &    $72^{+2}_{-2}$ &  $0.47^{+0.02}_{-0.02}$ &  $0.79^{+0.02}_{-0.03}$ &    $-54^{+5}_{-5}$ \\ SL2SJ1406$+$5226 &  0.716 &  $1.029^{+0.006}_{-0.006}$ &  $1.54^{+0.10}_{-0.12}$ &  $0.83^{+0.01}_{-0.01}$ &     $2^{+1}_{-1}$ &  $0.01^{+0.01}_{-0.01}$ &  $30^{+12}_{-11}$ &  $0.26^{+0.04}_{-0.04}$ &  $0.40^{+0.08}_{-0.05}$ &     $-6^{+5}_{-5}$ \\ SL2SJ2203$+$0205 &  0.400 &  $1.677^{+0.012}_{-0.012}$ &  $1.94^{+0.08}_{-0.08}$ &  $0.78^{+0.01}_{-0.01}$ &   $-50^{+1}_{-1}$ &  $0.09^{+0.01}_{-0.01}$ &    $39^{+1}_{-1}$ &  $0.54^{+0.01}_{-0.01}$ &  $0.77^{+0.01}_{-0.01}$ &    $-52^{+9}_{-9}$ \\ SL2SJ2205$+$0147 &  0.476 &  $1.814^{+0.042}_{-0.044}$ &  $1.92^{+0.13}_{-0.14}$ &  $0.35^{+0.02}_{-0.01}$ &   $-15^{+1}_{-1}$ &  $0.21^{+0.01}_{-0.01}$ &   $-18^{+2}_{-2}$ &  $0.43^{+0.02}_{-0.02}$ &  $0.49^{+0.03}_{-0.03}$ &      $0^{+2}_{-2}$ \\ SL2SJ2221$+$0115 &  0.325 &  $1.274^{+0.011}_{-0.011}$ &  $2.14^{+0.14}_{-0.11}$ &  $0.92^{+0.03}_{-0.02}$ &   $-10^{+6}_{-8}$ &  $0.06^{+0.01}_{-0.01}$ &   $-86^{+3}_{-2}$ &  $0.52^{+0.02}_{-0.02}$ &  $0.70^{+0.02}_{-0.02}$ &     $21^{+2}_{-2}$  \\ \hline
 \end{tabular}
\end{table*}

The lens model parameters are provided in Table~\ref{tab:results}. Image reconstruction and residuals for each system are shown in Appendix~\ref{appendix:models}. %We use the lensing parameters from these models to constrain the following CFHT model in Section~\ref{cfht}, as well as for the luminous and dark matter decomposition in Section~\ref{sec:hierarchial}.

\subsection{Modelling the lens galaxy light in the CFHT imaging}\label{cfht}

While the CFHT data has a lower resolution and $S/N$ compared to the \textit{HST} imaging, the CFHT \ipr-band is closer to the optical band absorption features used in the velocity dispersion measurements and accentuates the lens galaxy light over the source galaxy.  Hence, we use the CFHT \ipr-band imaging to fit the lens galaxy light profile of our SL2S systems. To accurately model the light profile, we deblend the lensed arcs in the CFHT data to robustly fit the lens galaxy's light distribution. We make use of the high constraining power of the lens models from Section \ref{sec:lens_modelling}, by imposing a posterior probability distribution (taken from the lens model) for the relative locations of the lens and source profiles and the mass model parameters for the CFHT \ipr-band models. %, in order to tightly constrain the lens light profile and reduce contamination from the source light.
\begin{figure*}
    \includegraphics[width=1\linewidth]{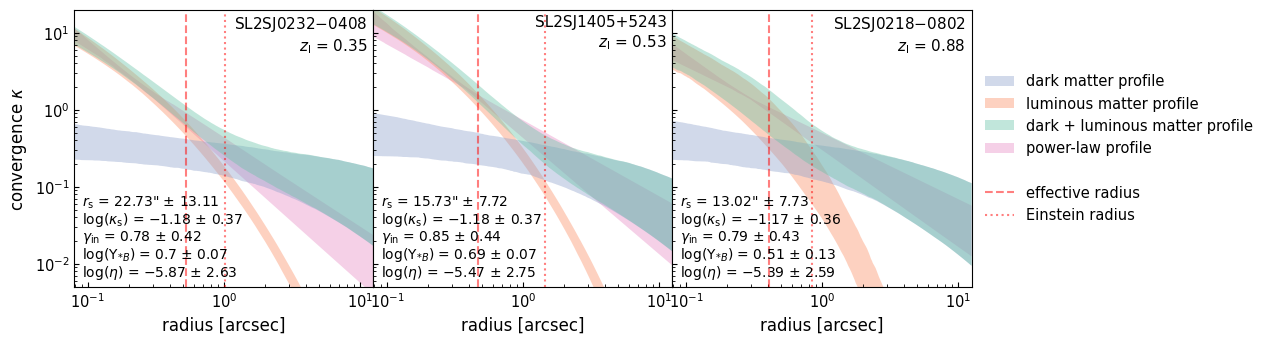}
    \caption{Luminous matter (LM) and dark matter (DM) decompositions for three strong lensing systems at varying redshifts in our sample.  We present the posterior distributions for each individual system using strong lensing information only, demonstrating its strong constraining power independent of kinematic data.  }%In the bottom row, we present the resulting population-level distributions (which account for kinematic and LOS information) imposed onto the same systems.  The darker-shaded regions correspond to the uncertainty in the mean $\mu$ population parameter, and the lighter-shaded regions also include the modelled intrinsic scatter $\sigma$.  The strong constraining power of the kinematic information allows for more accurate reconstruction of the inner radius matter distributions compared with lensing information alone.}
    \label{fig:example_decomp}
\end{figure*}
For the lens light profile, we use a single elliptical Sérsic profile with a uniform prior $n_{\rm s} \sim \mathcal{U}([1, 5])$ on the Sérsic index.  This is the same as the lens light model applied to the \textit{HST} imaging, but now with added flexibility in the Sérsic index.  We impose the same Gaussian prior on the lens light model centre and ellipticity as with the \textit{HST} data. For the source light profile, we also use a single elliptical Sérsic profile. This is similar to the source light model applied to the \textit{HST} imaging but without the shapelets basis, as we find that the CFHT~\ipr-band's comparatively lower angular resolution, redder colour, and lower $S/N$ in the lensed arcs do not require the additional freedom provided by shapelets.

The lensing model MCMC chains are then run until they have converged; see Table~\ref{tab:results} for the resulting lens light parameters.  The image reconstruction and residuals for each system are shown in Appendix~\ref{appendix:models}. Similar to the case of lens modelling with the \textit{HST} imaging, there are a few systems that do not strictly follow the model described above. We discuss these exceptional cases also in Appendix~\ref{appendix:models}. %We use the shape of the lens light profile from this model to represent an unscaled luminous matter density profile in our hierarchical analysis in Section~\ref{sec:hierarchial} (assuming a constant mass to light ratio).

\subsection{Estimating systematic uncertainty of the mass model parameters} \label{sec:systematic}

Only the statistical uncertainties are accounted for in the lens modelling procedures described above. While we expect some lensing quantities, such as the Einstein radius, to remain robust across modelling methodologies, $\gamma$ has been shown to vary significantly (Dinos-I).  To estimate systematic uncertainties in $\gamma$, a comparison between different models and modelling pipelines is necessary.  Of the 21 SL2S systems modelled in this paper, 11 overlap with Dinos-I's sample. Following Dinos-I, we use the lensing information quantity defined as
\begin{equation} \label{eq:lensing_info}
    \mathcal{I} = \frac{ \sum_{\rm i}^{\rm arc} w_i S_i}{\sqrt{ \sum_{\rm i}^{\rm arc} {N}_{i}^2}},
\end{equation}
where $S_i$ is the lens-light subtracted flux of the $i$th pixel on the lensed arcs, and $N_i$ is the noise level in that pixel.  The weight $w_i$ is defined as
% \begin{equation} \label{eq:lensing_info_weight}
%     w_i = \left[1 +  a \frac{\left| \theta_i - \theta_E \right|}{\theta_E} \left(1 + b \frac{\left| \phi_i - \phi_{\rm ref} \right| }{\phi_{\rm ref}} \right)  \right]
% \end{equation}
%
\begin{equation} \label{eq:lensing_info_weight}
    w_i = \left[1 +  \frac{\left| \theta_i - R_{\rm E} \right|}{R_{\rm E}} \left(1 + \frac{\left| \phi_i - \phi_{\rm ref} \right| }{\phi_{\rm ref}} \right)^b  \right]^{a},
\end{equation}
where $\theta_{i}$ is the radial distance of the pixel from the deflector galaxy, $R_{\rm E}$ is the Einstein radius, $\phi_i$ is the azimuthal angle of the pixel, and $\phi_{\rm ref}$ is a reference angle which we chose to correspond to the brightest pixel on the arc.  By minimizing the correlation between $\log \mathcal{I}$ and $\log \sigma_{\gamma}$ (where $\sigma_{\gamma}$ is the statistical uncertainty of $\gamma$) across our 21 SL2S systems and Dinos-I's 77 systems, we estimate $a \approx 2$ and $b \approx 0.2$.  

Following Dinos-I (see Figure~7 therein),  the systematic uncertainty given a system's lensing information is described with
\begin{equation} \label{eq:gamma_sys_uncertainty}
    \sigma_{\gamma, \rm sys} = \sigma_{\gamma, \rm sys}^{\rm max} \tanh \left( {\mathcal{I}_{\rm scale}} / \mathcal{I} \right).
\end{equation}
Dinos-I sets $\sigma_{\gamma, \rm sys}^{\rm max} = 1/3$, which we also adopt in this paper.  $\mathcal{I}_{\rm scale}$ is fit for by minimizing the absolute difference between unity and the reduced $\chi^2$, summed over the 11 overlapping systems between our and Dinos-I's samples, that is, this penalty function defined as
\begin{equation} \label{eq:gamma_sys_i_scale}
    \mathcal{P}(\mathcal{I}_{\rm scale}) \equiv  \left| 1 - \frac{1}{N_{\rm lens}} \sum_{j}^{N_{\rm lens}} \frac{\left(\gamma^{\rm ours}_j - \gamma^{\rm Tan+24}_j \right)^2}{(\sigma^{\rm ours}_{\gamma,j})^2 + (\sigma^{\rm Tan+24}_{\gamma,j})^2} \right|.
\end{equation}
From this, we find $\mathcal{I}_{\rm scale}=125$.  The total uncertainty is then calculated by adding the systematic and statistical uncertainties in quadrature.  See Figure~\ref{fig:gamma_systematics} for a summary comparison between our and Dinos-I's $\gamma$ values, after accounting for systematic uncertainties.  Note that since we fit for different weights $a$ and $b$, the information criterion $\mathcal{I}$ values we present differ from those presented in Dinos-I. Figure~\ref{fig:gamma_systematics} (left panel) shows that the results are not only consistent with Dinos-I, but also have smaller uncertainties as the F475X data provides higher $S/N$ within the lensing features (see Section~\ref{sec:observation}, and Figures~\ref{fig:sl2s_comparision0}, \ref{fig:sl2s_comparision1}, and \ref{fig:sl2s_comparision2} for a qualitative comparison).  The same methodology is applied to estimate our sample's Einstein radii systematic uncertainties as well. 
On average, the $\gamma$ systematic uncertainty makes the total uncertainty 5.7 times larger than the statistical uncertainty, while the $R_{\rm E}$ systematic uncertainty does so by a factor of 2.7.  The total uncertainties are presented in Table~\ref{tab:results}.

\section{Hierarchical analysis} \label{sec:hierarchial}
Our goal is to constrain the population-level properties of the luminous and dark matter profiles for our full sample of SLACS and SL2S lenses. To that aim, we perform a hierarchical Bayesian analysis on our sample. However, to run in a computationally reasonable time, we first parameterize each system at an individual level before constraining the associated population-level parameters. In this paper, most of the population-level parameters are denoted as such: $\mu_x$ is the population mean and $\sigma_x$ is the population scatter of the subscripted parameter $x$.  If we account for a redshift evolution in $x$, then $\alpha_x$ is the linear correlation for that parameter with respect to redshift $z$, $\epsilon_x$ is the linear correlation for that parameter with respect to the logarithmic velocity dispersion $\log(\sigma_{\rm v})$, and $\mu_x$ and $\sigma_x$ are evaluated at the reference redshift $z_{\rm ref} \equiv 0.353$ and reference velocity dispersion $\log(\sigma_{\rm v, ref}) \equiv 2.42$, which are the means of our 56 systems sample.

We adopt a generalized NFW profile \citep[gNFW;][]{keeton2001, wyithe2001} for modelling the dark matter profile, allowing us to quantify how the inner slope of these halos evolve over redshift due to baryonic processes. The gNFW density profile is defined as 
\begin{equation}\label{gnfw_eq}
\rho (r) = \frac{\rho_{\rm s}}{(r/r_{\rm s})^{\gamma_{\rm in}}(1+r/r_{\rm s})^{3-\gamma_{\rm in}}},
\end{equation}
where $r_{\rm s}$ is the scale radius, $\rho_{\rm s}$ is the density at $r_{\rm s}$, and $\gamma_{\rm in}$ is the inner logarithmic slope.  With $\gamma_{\rm in} = 1$, the gNFW profile becomes an NFW profile.  This parameter $\gamma_{\rm in}$ and its evolutionary trend can provide insights into the baryonic processes that have shaped the dark matter and generally the overall mass distribution (mergers, stellar, supernovae, and AGN feedback, etc.), with $\gamma_{\rm in} < 1$ roughly describing an expanded or cored halo \citep[e.g.,][]{Governato2010, Governato2012} and $\gamma_{\rm in} > 1$ roughly describing a contracted or cuspier halo \citep[e.g.,][]{gnedin2011, newman2013}.  

As we are interested in constraining the population-level stellar mass-to-light ratio ($\Upsilon_{*B} \equiv M_*/L_B$, where both are in solar units), it is necessary to convert the modelled surface brightness, i.e., the S\'ersic profile(s), to a $B$-band luminosity scale.  For our modelled SL2S systems, we use the CFHT \ipr-band light profiles (see Section~\ref{cfht}).  For the remaining SL2S and SLACS systems, we use the \textit{HST} light profiles modelled by Dinos-I.  As such, all systems' S\'ersic profile amplitudes were scaled according to their distance modulus (assuming our fiducial cosmology) and $K$-corrected to correspond to the rest-frame $B$-band, requiring SED fitting.  We use \textsc{kcorrect}\footnote{\url{https://github.com/blanton144/kcorrect}}, a \textsc{python} implementation for calculating these $K$-corrections  \citep{blanton2007}, to account for the distance modulus (the angular diameter distance and cosmological surface brightness dimming corrections) and the SED shape variation.  For the SL2S systems, we use simple aperture photometry (of radius $\sim0.1$ arcsec, centred on the lens) on the CFHT $u$, $g$, $r$, $i$, and $z$-bands for the necessary colour information; for the SLACS systems, we use the colour information of \citet{auger2009} (from \textit{HST} F435W, F555W, F606W, F814W, and/or F160W filters).  The resulting distance-modulus-corrected and $K$-corrected surface brightness profile in $B$-band is denoted as $I_{B}(r) $. %\michele[Can you explicitly state and make sure that kcorrect accounts for (i) the SED shape variation between the two bands, (ii) the $(1+z)$ factor due to the redshifting of the band in the AB system and (iii) the $(1+z)^{-4}$ cosmological bolometric surface brightness dimming. This is to reassure the reader that you are aware of the issues.]{

%\michele[This parameter $\eta$ is completely degenerate with $\gamma_{in}$. We need to make this clear. The only usefulness of fitting for both is to have more realistic uncertainties on $\eta$, not to measure both. The lensing and dynamics only measure the total density profile. Any change in $\gamma_{in}$ can be compensated by a change in $\eta$ to keep the total density constant. If you fit for both, your result is driven by the priors and subtle assumptions in your model (the effect of $\eta$ is not identical to $\gamma_{in}$).]{In addition to $\Upsilon_{*B}$, we also include a population stellar mass-to-light gradient component $\eta$.}  
In addition to $\Upsilon_{*B}$, we also include a population stellar mass-to-light gradient component $\eta$.
This is incorporated by multiplying the luminous matter surface density by a power law dictated by $-\eta$. Quantitatively, therefore, the luminous matter convergence $\kappa_{*}$ is modelled as
\begin{equation} \label{eq:lum_kappa}
\kappa_{*}(r) = \frac{\Upsilon_{*B}}{\Sigma_{\rm crit}} \left( \frac{r}{R_{\rm eff}} \right) ^{- \eta} I_{B}(r)
\end{equation}
where $\Sigma_{\rm crit}$ is the lensing critical surface mass density. Although lensing and dynamics constrain the radial shape of the overall mass density profile, leaving a degeneracy between the dark and luminous matter distribution in the most general sense, varying $\eta$ and $\gamma_{\rm in}$ affects the overall mass profile shape differently in our particular parametrization. As a result, we do not have a perfect degeneracy between $\eta$ and $\gamma_{\rm in}$, as it is broken by our physically motivated model parametrization. Furthermore, we find including $\Upsilon_{*B}$ in our model makes the best-fit anisotropy profile more consistent with previous empirical studies (Section~\ref{sec:results}).

We include a $\Upsilon_{*B}$ prior to better constrain our model.  For the redshift evolution parameter $d\log(\Upsilon_{*B})/dz$, we adopt a prior of $\alpha_{\log(\Upsilon_{*B})} = -0.72^{+0.07}_{-0.05} \text{ (statistical)} \pm 0.04 \text{ (systematic)}$ \citep{treu2005}.
%, which is close to the independent result by \citet{vanderMarel2007}.  
This prior is derived from fundamental plane analysis, a methodology independent of lensing, on spheroidal galaxies.  For the corresponding normalization, we analyze a set of 79 SLACS lenses from \citet{auger2009}, as they share a similar velocity dispersion distribution to our sample (also considering the nonoverlapping subsample of SL2S lenses), and their constraints on the stellar profiles are independent of lensing analysis.  We calculate $\log(\Upsilon_{*B})$ for each of the 79 systems using the stellar mass and $B$-band absolute magnitudes provided in \citet{auger2009}.  We obtain the population mean of $\Upsilon_{*B}$ and its uncertainty from these 79 systems to serve as the prior for our sample. In this sampling, we set the prior for the stellar initial mass function (IMF) to be uniform between the Salpeter and Chabrier IMF to make the prior uninformative on the IMF. In practice, we obtain $\log(\Upsilon_{*B})$ from \citet{auger2009} assuming a Salpeter IMF \citep{salpeter}.  Then with an additional offset factor of $\mathcal{U}([-\log(1.8), 0])$ to uniformly sample between the Chabrier \citep{chabrier} and Salpeter IMFs, we sample the $\mu_{\log(\Upsilon_{*B})}$ posterior and fit the distribution as a Gaussian; in contrast, the intrinsic scatter $\sigma_{\log(\Upsilon_{*B})}$ is measured at a fixed IMF. We also sample from the aforementioned $\alpha_{\log(\Upsilon_{*B})}$ prior, to fit for $\mu_{\log(\Upsilon_{*B})}$ and $\sigma_{\log(\Upsilon_{*B})}$ measured at $z_{\rm ref}$. 
From this, we find and use the following as our prior:
\begin{equation} \label{eq:mlprior0}
\begin{matrix*}[l]
    \mu_{\log(\Upsilon_{*B})} \sim \mathcal{N}( \mu=0.46, \sigma=0.08),\\
    \sigma_{\log(\Upsilon_{*B})} \sim \ln\mathcal{N}(\mu=-3.94, \sigma=1.18),\\
    \alpha_{\log(\Upsilon_{*B})} \sim \mathcal{N}(\mu=-0.72, \sigma=0.07),
\end{matrix*}
\end{equation}
where we define a lognormal distribution as
\begin{equation} \label{eq:lognormal}
\ln\mathcal{N}(\mu, \sigma) \equiv e^{\mathcal{N}(\mu, \sigma)} .
\end{equation}

We first calculate the posterior distributions for the individual-level parameters pertaining to the luminous and dark matter profiles for each system, before performing the analysis at the population level.  Equation~(\ref{eq:lum_kappa}) is used to 
%\michele[Can you state somewhere that the dynamical models make a distinction between the tracer population, for which we adopt the surface brightness $I_B(r)$, and the mass density, for which we consider the $M/L$ gradients of eq.(8). This has been done incorrectly in some past papers (not by the co-authors).]
convert the lens light surface luminosity to luminous matter surface density used in our dynamical models; the overall convergence shape is set by the stellar mass-to-light ratio parameter $\Upsilon_{*B}$, the stellar mass-to-light gradient parameter $\eta$, and the 
corrected, modelled $B$-band surface luminosity profile $I_B(r)$.  We define $D_{\rm light}$ as the combined dataset of $I_B(r)$ across all systems.

As we use the gNFW model for the dark matter profile, for each system of our sample, we obtain posterior distributions of the individual-level parameters $\tau \equiv \{ r_{\rm s}, \log(\kappa_{\rm s}), \gamma_{\rm in}, \log(\Upsilon_{*B}), \eta \}$, where $\kappa_{\rm s}$ is the convergence of the gNFW profile at $r_{\rm s}$ (i.e., $\kappa_{\rm s} \equiv \rho_{\rm s}r_{\rm s}/\Sigma_{\rm crit}$; see equation~\ref{gnfw_eq}).  

At the individual-system level, we constrain over only the lensing observables $D_\kappa$ independent of the mass-sheet degeneracy: the Einstein radius and the quantity $R_{\rm E} \alpha^{\prime \prime}_{\rm E} / (1- \kappa_{\rm E})$, where $\kappa_{\rm E}$ and $\alpha_{\rm E}$ are the convergence and deflection angle at the Einstein radius, respectively \citep{kochanek2020, Birrer2021}.  

Additionally, we impose a joint prior on $R_{\rm s}$ and $\log(\kappa_{\rm s})$, in the form of a mass--concentration relation for the dark matter profile \citep{duffy2008}\txr{, which was derived assuming a WMAP5 cosmology \citep{Komatsu2009}. 
 Given that we utilize a gNFW dark matter profile, we adopt a concentration definition of 
\begin{equation} \label{eq:gNFW_concentration}
    c_{-2} \equiv r_{200} / r_{-2},
\end{equation}
where $c_{-2}$ is the modified gNFW concentration parameter, $r_{200}$ is the radius where the mean density is 200 times the critical density of the universe, and $r_{-2}$ is the radius at which the logarithmic surface density slope is equal to $-2$ \citep{keeton2001}.  In the case of an NFW profile ($\gamma_{\rm in} = 1$), $c_{-2}$ is equivalent to the conventional concentration parameter of $r_{200}/r_{\rm s}$.  While the mass--concentration relation from \citet{duffy2008} was estimated using NFW dark matter simulated halos, we assume that this relation also applies to $c_{-2}$, as they are both similarly-physically motivated \citep{keeton2001}. }

The $\log(\Upsilon_{*B})$ parameter is sampled from a broad uniform distribution encompassing $3\sigma$ level of population scatter for $\mu_{\log(\Upsilon_{*B})}$ in equation~(\ref{eq:mlprior0}) scaled by their respective redshift, that is, $\mathcal{U}([0.21, 0.71]) \times -0.72(z - 0.347)$. We take this uniform distribution for the prior for individual systems as the values are informed by the population level prior and, therefore, we allow more freedom in the prior for individual systems than the posterior of the mean would allow. 

Furthermore, $\gamma_{\rm in}$ and $
\eta$ also have priors of $\mathcal{U}([0.1, 2.0))$ and $\mathcal{U}(\ln((0, 0.5]))$, respectively.  The resulting individual posteriors are parameterized as Gaussian distributions, with exception to $\eta$, which is parameterized as a lognormal distribution.  \txr{ This is because we know $\eta$ cannot be $<0$ based on studies of elliptical galaxies of stellar populations and dynamics \citep{Lu2024} and IMF gradients \citep{vandokkum17}}.  See Appendix~\ref{appendix:decomps} for the individual posteriors for each system in our sample.  Therefore, the posterior probability density function $p(\tau_i \mid D_{\kappa, i}, D_{{\rm light}, i})$ for an individual lens is
\begin{equation} \label{eq:individualposterior}
    p(\tau_i \mid D_{\kappa, i}, D_{{\rm light}, i}) \propto p(\tau_i)\, p(D_{\kappa, i} \mid \tau_i)\, p(D_{{\rm light}, i} \mid \tau_i), 
\end{equation}
where $p(\tau_i)$ is the combination of aforementioned priors, $p(D_{\kappa, i} \mid \tau_i)$ is the likelihood of the lensing observables, and $ p(D_{{\rm light}, i} \mid \tau_i)$ is the likelihood of the observed surface brightness profiles.

In Figure~\ref{fig:example_decomp}, we show the luminous and dark matter distributions of three exemplar systems in our sample (\mbox{SL2SJ0232$-$0408}, \mbox{SL2SJ0226$-$0420}, and \mbox{SL2SJ018$-$0802}) at varying redshifts, and how their decomposition compares to the modelled power-law density profile.  The luminous and dark matter decomposition figures for our full sample of 56 systems are presented in  Appendix~\ref{appendix:decomps}.

\begin{table*} 
 \caption{Description and priors of the population-level parameters sampled by our Bayesian hierarchical process.  See Section \ref{sec:hierarchial} for the derivation of our priors on $\mu_{\log(\Upsilon_{*B})}$, $\sigma_{\log(\Upsilon_{*B})}$, and $\alpha_{\log(\Upsilon_{*B})}$.}
 \label{tab:pop_names}
 \begin{tabular}{lllll}
  \hline
  Parameter & Description & Prior %& \multicolumn{1}{|p{1cm}|}{\centering Posterior\\ (constant anisotropy)} & \multicolumn{1}{|p{1cm}|}{\centering Posterior (OM anisotropy)}
  \\
  \hline
$\mu_{\gamma_{\rm in}}$ & Population mean of $\gamma_{\rm in}$, the logarithmic inner slope of the DM halo, at $z_{\rm ref}$ and $\log(\sigma_{\rm v, ref})$ & $\mathcal{U}([0.1, 2.0))$ \\ 
$\sigma_{\gamma_{\rm in}}$ & Population scatter of $\gamma_{\rm in}$ at $z_{\rm ref}$ and $\log(\sigma_{\rm v, ref})$ & $\mathcal{U}([0, 1])$ \\
$\alpha_{\gamma_{\rm in}}$ & Linear dependency of $\gamma_{\rm in}$ on redshift & $\mathcal{U}([-5, 5])$ \\ 
$\epsilon_{\gamma_{\rm in}}$ & Linear dependency of $\gamma_{\rm in}$ on $\log(\sigma_{\rm v})$& $\mathcal{U}([-5, 5])$ \\
$\mu_{\log(\Upsilon_{*B})}$ & Population mean of $\log(\Upsilon_{*B})$, the log stellar mass-to-$B$-band-light ratio, at $z_{\rm ref}$ and $\log(\sigma_{\rm v, ref})$ & $\mathcal{N}(\mu=0.46, \sigma=0.08)$ \\
$\sigma_{\log(\Upsilon_{*B})}$ & Population scatter of $\log(\Upsilon_{*B})$ at $z_{\rm ref}$ and $\log(\sigma_{\rm v, ref})$ & $\ln\mathcal{N}(\mu=-3.94, \sigma=1.18)$ \\
$\alpha_{\log(\Upsilon_{*B})}$ & Linear dependency of $\log(\Upsilon_{*B})$ on redshift & $\mathcal{N}(\mu=-0.72, \sigma=0.07)$ \\
$\epsilon_{\log(\Upsilon_{*B})}$ & Linear dependency of $\log(\Upsilon_{*B})$ on $\log(\sigma_{\rm v})$& $\mathcal{U}([-5, 5])$ \\
$\overline{\eta}$ & Population average stellar mass-to-light gradient & $\mathcal{U}([0, 0.5])$ \\
$\mu_{\beta_{\rm ani}}$ & Population mean of $\beta_{\rm ani}$, the stellar anisotropy parameter & $\mathcal{U}([-1, 1])$ \\
$\sigma_{\beta_{\rm ani}}$ & Population scatter of $\beta_{\rm ani}$ & $\mathcal{U}(\ln((0, 0.5]))$ \\
$\mu_{a_{\rm ani}}$ & Population mean of $a_{\rm ani}$, the anisotropy radius divided by the effective radius & $\mathcal{U}(\ln([0.2, 5]))$ \\
$\sigma_{a_{\rm ani}}$ & Population scatter of $a_{\rm ani}$ & $\mathcal{U}(\ln((0, 0.5]))$ \\
% $\zeta_{{\rm SDSS}, {\rm sys}}$ & Fractional systematic uncertainty for SDSS $\sigma_{\rm v}$ measurements (see equation~(\ref{eq:vd_systematics})) & $\mathcal{U}(\ln((0, 0.5]))$ \\
\hline
\end{tabular}
\end{table*}
% \addtolength{\tabcolsep}{-0.1em}
\begin{table*} 
 \caption{1D posteriors of the population-level parameters sampled by our Bayesian hierarchical process.  The baseline model assumes a constant anisotropy, the priors established in Table~\ref{tab:pop_names}, $H_0 = 70$ km s$^{-1}$ Mpc$^{-1}$, and is applied to our full 56 SL2S and SLACS lens sample.  The 95th percentile value is provided for the population-level scatter parameters and $\overline{\eta}$.}  %As we do not utilize SDSS velocity dispersion measurements for the SL2S systems, $\zeta_{{\rm SDSS}, {\rm sys}}$ is not applicable to the SL2S-only model. }
 \label{tab:posteriors}
 \begin{tabular}{cccccccccccc}
  \hline
  Model & $\mu_{\gamma_{\rm in}}$ & $\sigma_{\gamma_{\rm in}}$ & $\alpha_{\gamma_{\rm in}}$ & $\epsilon_{\gamma_{\rm in}}$ & $\mu_{\log(\Upsilon_{*B})}$ & $\sigma_{\log(\Upsilon_{*B})}$ & $\alpha_{\log(\Upsilon_{*B})}$ & $\epsilon_{\log(\Upsilon_{*B})}$ &
  $\overline{\eta} \times 10^{3}$ & 
  $\mu_{\beta_{\rm ani}}$ or $\mu_{a_{\rm ani}}$ & $\sigma_{\beta_{\rm ani}}$ or $\sigma_{a_{\rm ani}}$ 
  % $\zeta_{{\rm SDSS}, {\rm sys}}$ 
  \\
  \hline

Baseline & $0.97^{+0.03}_{-0.03}$ & $\le0.07$ & $-0.44^{+0.14}_{-0.15}$ & $0.39^{+0.39}_{-0.41}$ & $0.72^{+0.02}_{-0.02}$ & $\le0.19$ & $-0.57^{+0.06}_{-0.06}$ & $0.56^{+0.28}_{-0.28}$ & $\le0.05$ & $-0.20^{+0.21}_{-0.19}$ & $\le0.46$ \\
OM & $0.95^{+0.04}_{-0.03}$ & $\le0.08$ & $-0.48^{+0.15}_{-0.16}$ & $0.47^{+0.37}_{-0.44}$ & $0.72^{+0.03}_{-0.02}$ & $\le0.20$ & $-0.57^{+0.07}_{-0.08}$ & $0.63^{+0.36}_{-0.35}$ & $\le0.05$ & $3.62^{+0.91}_{-1.36}$ & $\le0.44$ \\ \hline
SL2S & $0.92^{+0.14}_{-0.15}$ & $\le0.28$ & $-0.18^{+0.43}_{-0.41}$ & $0.62^{+0.88}_{-1.06}$ & $0.75^{+0.05}_{-0.08}$ & $\le0.36$ & $-0.67^{+0.09}_{-0.10}$ & $0.32^{+0.54}_{-0.59}$ & $\le1.89$ & $-0.19^{+0.19}_{-0.19}$ & $\le0.36$ \\
SLACS & $0.95^{+0.07}_{-0.06}$ & $\le0.08$ & $-0.46^{+0.43}_{-0.38}$ & $0.09^{+0.55}_{-0.63}$ & $0.62^{+0.02}_{-0.02}$ & $\le0.14$ & $-0.74^{+0.07}_{-0.07}$ & $1.06^{+0.46}_{-0.39}$ & $\le0.13$ & $-0.04^{+0.19}_{-0.22}$ & $\le0.45$ \\ \hline
$H_0 = 67$ & $0.97^{+0.03}_{-0.03}$ & $\le0.07$ & $-0.45^{+0.16}_{-0.14}$ & $0.41^{+0.41}_{-0.36}$ & $0.73^{+0.02}_{-0.02}$ & $\le0.19$ & $-0.58^{+0.06}_{-0.05}$ & $0.60^{+0.36}_{-0.43}$ & $\le0.05$ & $-0.20^{+0.21}_{-0.19}$ & $\le0.45$ \\
$H_0 = 73$ & $0.96^{+0.03}_{-0.03}$ & $\le0.07$ & $-0.46^{+0.13}_{-0.14}$ & $0.41^{+0.38}_{-0.36}$ & $0.73^{+0.02}_{-0.02}$ & $\le0.19$ & $-0.58^{+0.07}_{-0.07}$ & $0.59^{+0.30}_{-0.29}$ & $\le0.05$ & $-0.19^{+0.20}_{-0.17}$ & $\le0.47$
\\

\hline
 \end{tabular}
\end{table*}

\begin{figure*}
    \includegraphics[width=1\linewidth]{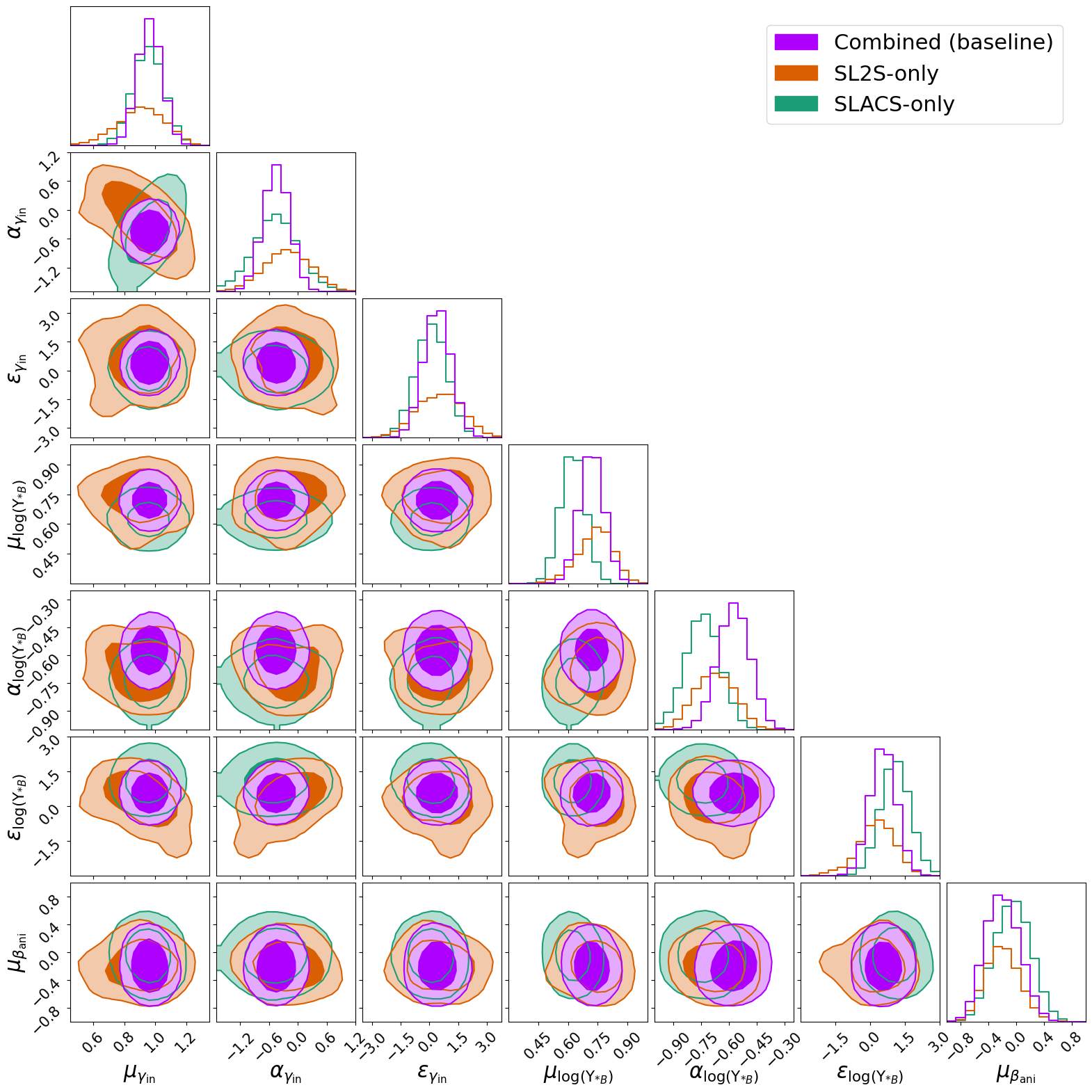}
    \caption{Posteriors of the population-level parameters from our Bayesian hierarchical analysis applied on our full sample (purple), SL2S-only (orange), or SLACS-only (green).  See Table~\ref{tab:pop_names} for a description of all the parameters and their marginalized point estimates.  For the sake of readability, we exclude the population-level scatter and $\overline{\eta}$ parameters from this figure, though their 95th percentile values are given in Table~\ref{tab:pop_names}.  The darker and lighter shaded regions in the 2D distributions represent 68th and 95th percentiles, respectively.  All model parameters shown qualitatively seem to follow well a Gaussian distribution.}\label{fig:corner}
\end{figure*}

\begin{figure*}
    \includegraphics[width=0.8\linewidth]{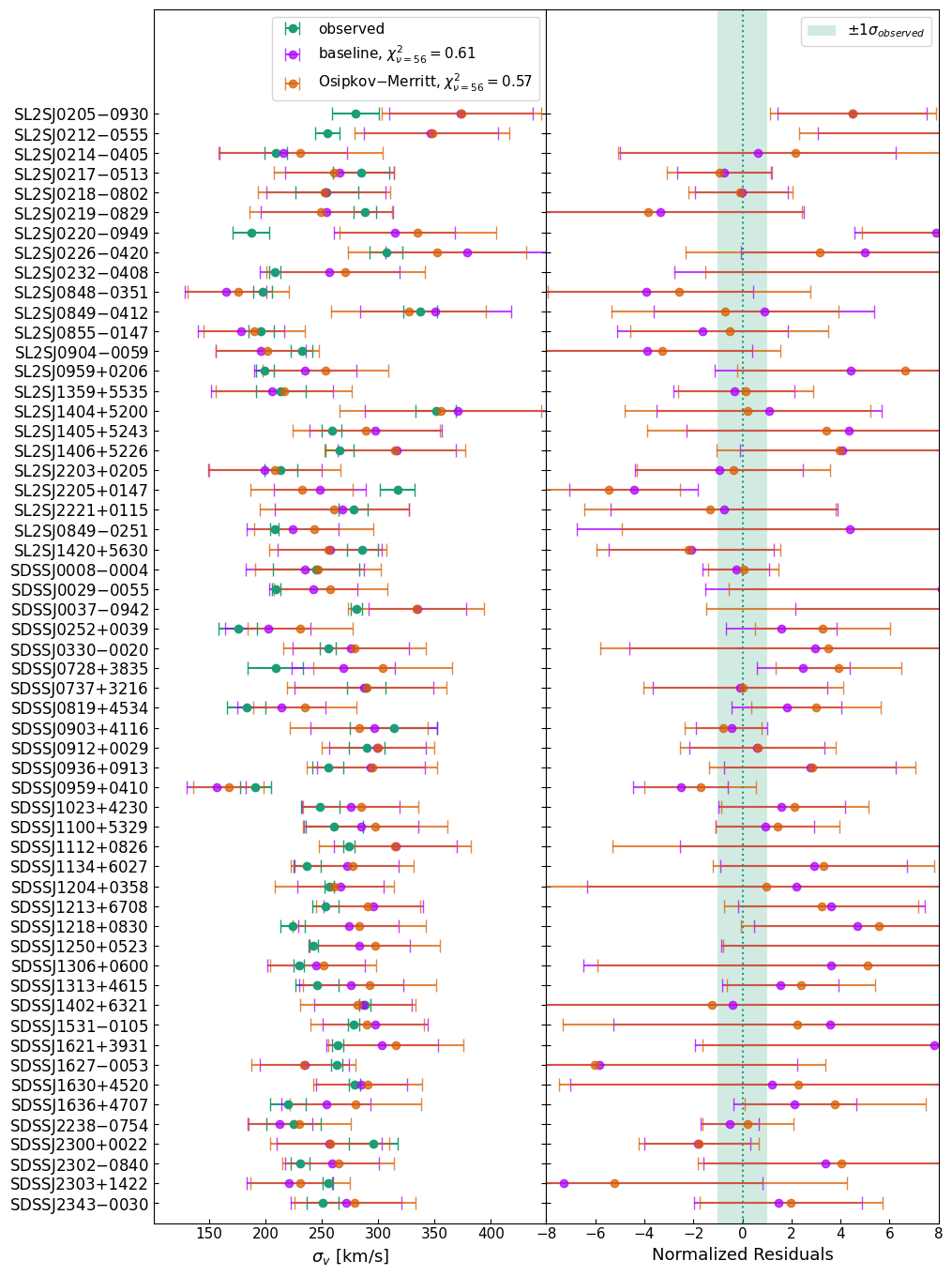}
    \caption{Comparison between the observed and predicted velocity dispersions (left) and their normalized residuals (right, normalized with the measurement uncertainties) from our Bayesian hierarchical analysis, for all 56 SLACS and SL2S lenses in our sample.  
    % For the measured SDSS measurements (i.e., the measured velocity dispersions for the SLACS systems), $\zeta_{{\rm SDSS}, {\rm sys}} = 0.13$ is applied for the systematic uncertainty (see equation~(\ref{eq:vd_systematics}) and Table~\ref{tab:posteriors}).  
    We show the predictions for spatially constant (orange) and Osipkov--Merritt (green) anisotropy profiles; their respective reduced $\chi^2_{\nu}$ (solely based on the velocity dispersion measurements) are given in the legend.  This plot shows the consistency between observed and measured velocity dispersion, across both anisotropy models.}
    \label{fig:vd}
\end{figure*}

\begin{figure}
    \includegraphics[width=1\linewidth]{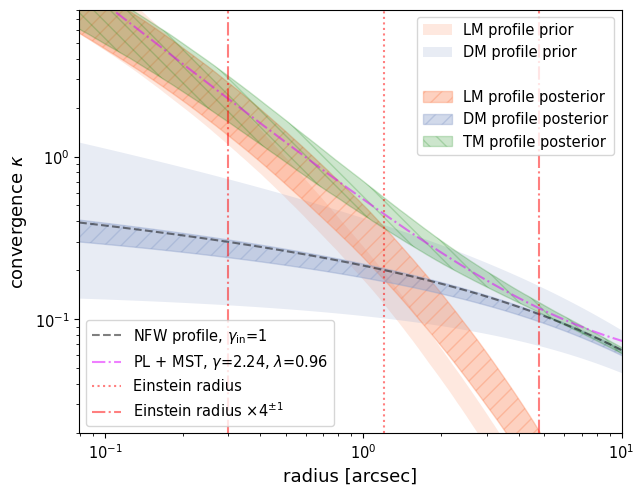}
    \caption{The luminous (LM), dark (DM), and total matter (TM) profiles of a representative, averaged system for our lens sample (see text for details) at $z_{\rm ref}$ and $\log(\sigma_{\rm v, ref})$.  The light and dark-shaded regions represent the prior (without kinematic and LOS information) and the hierarchical posterior distributions (with kinematic and LOS information), respectively, for a given coloured profile.  The dark matter NFW curve ($\gamma_{\rm in} = 1$) is shown in grey. 
 The total density profile posterior is plotted as the green shaded region, and the best-fitting power law + MST model plotted in purple.  Within the relevant lensing regime of $0.25 R_{\rm E}$ to $4 R_{\rm E}$ (red dotted-dashed lines), we see that the total density profile agrees well with a power law profile with a small contribution from the MST ($\lambda \approx 1$).}% is consistent with the total matter profile between $0.25 R_{\rm E}$ and $4 R_{\rm E}$.}
    \label{fig:representative}
\end{figure}

Finally, we use \textsc{HierArc}\footnote{\url{https://github.com/sibirrer/hierarc}} \citep[][]{birrer2020}, which is a hierarchical Bayesian inference pipeline for strong lensing systems, to infer the population-level parameters $\pi \equiv \{ \mu_{\gamma_{\rm in}}, \sigma_{\gamma_{\rm in}}, \alpha_{\gamma_{\rm in}}, \epsilon_{\gamma_{\rm in}}, \mu_{\log(\Upsilon_{*B})}, \sigma_{\log(\Upsilon_{*B})}, \alpha_{\log(\Upsilon_{*B})}, \epsilon_{\log(\Upsilon_{*B})}, \overline{\eta}, \mu_{\beta_{\rm ani}}, \\
\sigma_{\beta_{\rm ani}} \}$.  See Table~\ref{tab:pop_names} for a description of the parameters.  Unlike with the other population parameters, $\overline{\eta}$ is parameterized as the population-averaged $M_*/L_B$ gradient; this is equivalent to a population mean parameterization of the gradient with a scatter of zero, or a Dirac delta function. 
%\michele[See previous comment: I am unconvinced we can asses this]{This is because we are interested in probing whether a $\eta$ is necessary for our population of lenses (see Section~\ref{subsec:cosmography}), not to study the significantly-unconstrained intrinsic scatter.}
% We incorporate the single-aperture velocity dispersion measurements for all 58 systems.  For the SLACS systems with velocity dispersion measurements obtained via SDSS observations, we also account for a fractional systematic uncertainty
% %
% \begin{equation} \label{eq:vd_systematics}
%     \zeta_{\rm SDSS, sys} = \frac{\sigma_{\sigma_{\rm v}, {\rm sys}}}{\sigma_{\rm v}},
% \end{equation}
% %
% where $\sigma_{\sigma_{\rm v}, {\rm sys}}$ is the systematic uncertainty of SDSS $\sigma_{\rm v}$ measurements.  We implement this due to observational evidence that the SDSS uncertainties are significantly underestimated (\citealt {birrer2020}; Knabel et al. 2024, in preparation).  
We also estimate external convergence distributions for all but five systems (for which we do not have direct estimates of the external convergence) into this hierarchical analysis.  For the remaining five SL2S/SLACS systems, we apply the combined $\kappa_{\rm ext}$ distribution from the other SL2S/SLACS systems, respectively.

In addition to probing galaxy properties' relation versus redshift, we also account for their linear relation with $\log(\sigma_{\rm v})$ (i.e., $\epsilon_{\gamma_{\rm in}}$ and $\epsilon_{\log(\Upsilon_{*B})}$).  The primary reason for this is so that our results remain unbiased against the selection bias that is present in the SLACS sample.  As \citet{Sonnenfeld2024} shows, the SLACS lenses have steeper density profile and larger velocity dispersion than regular galaxies, at fixed stellar mass; but they are near indistinguishable at fixed velocity dispersion.  While we acknowledge that there are bound to be other differences between the SLACS and SL2S selection of lenses, we account for the seemingly most significant differences (redshift and velocity dispersion).  Hence, we are able to effectively combine the SLACS and SL2S samples, with the results being applicable to elliptical galaxy populations.  

As the baseline setting for the anisotropy profile, we assume a spatially constant anisotropy model, which has been shown to be consistent with local elliptical galaxies \citep{gerhard2001, cappellari2007, Cappellari16}. In this model, the stellar anisotropy parameter $\beta_{\rm ani}$ is spatially constant, which is defined as
\begin{equation} \label{eq:constant_ani}
    \beta_{\rm ani} = 1-\frac{\sigma_{\rm t}^2}{\sigma_{\rm r}^2} ,
\end{equation}
where $\sigma_{\rm t}$ and $\sigma_{\rm r}$ are the tangential and radial velocity dispersions, respectively.

For comparison with previous work, we also implement an Osipkov--Merritt anisotropy model \citep{Osipkov, Merritt85} in addition to a constant stellar anisotropy.  Under this model, we parameterize $a_{\rm ani}$, which is defined as the anisotropy radius divided by the effective radius \citep{mamon2005}:
% \michele[I would give a sentence about how you solve the Jeans equations: I understand LENSTRONOMY solves eq.(A15) of Mamon+05, using the Sersic profile for the light and your eq.(5)--(6) for the mass. What about deprojection of eq.(6) and integration for $M(r)$? Are you using MGE and eq.(50) of spherical JAM (Cappellari 2008)? In that case you may state it. Then you convolve with the PSF also numerically. How did you measure the PSF?]
\begin{equation} \label{eq:om_ani}
    a_{\rm ani} = r_{\rm ani} / r_{\rm eff} ,
\end{equation}
with the spatially-varying stellar anisotropy defined as
\begin{equation} \label{eq:om_ani_beta}
    \beta_{\rm ani, OM}(r) = \frac{r^2}{r^2 + a_{\rm ani}^2 r_{\rm eff}^2} .
\end{equation}
We sample from the parent population (as defined by population-level parameters $\pi$), to get $\xi \equiv \{ \gamma_{\rm in}, \log(\Upsilon_{*B}), \eta, \beta_{\rm ani} \}$ for each system.  The model-predicted velocity dispersion is then determined by deprojecting the total-matter 2D convergence and surface light profiles through a multi-Gaussian expansion process \citep{Emsellem1994, Cappellari2002}, where the 2D profile is decomposed into a Gaussian basis then individually deprojected into the corresponding 3D profile.  

Therefore, the population-level posterior probability density function $p(\pi \mid D)$ is
%We sample from the $R_s$ and $\log(\kappa_s)$ individual distributions, while marginalizing over the $\gamma_{\rm in}$ distributions and velocity dispersion measurements for each system.  
\begin{equation} \label{eq:popposterior1}
    p(\pi \mid D) \propto p(\pi) 
\prod_i^N \int \,d \xi_i p(\xi_i \mid \pi) p(D_i \mid \xi_i) ,
\end{equation}
where $D$ is the full set of data for all systems, and $D_i$ is the data pertaining to a single system.  Decomposing $D_i \equiv \{D_{\sigma_v, i}, D_{{\rm LOS}, i}, D_{\kappa, i}, D_{{\rm light}, i} \}$ (where $D_{\sigma_v}$ refers to the observed velocity dispersions and $D_{{\rm LOS}, i}$ refers to the external convergence measurements) and combining with equation~(\ref{eq:individualposterior}), we see that
\begin{multline} \label{eq:popposterior2}
    p(\pi \mid D) \propto p(\pi) 
\prod_i^N \biggl[ \int \,d \xi_i \,d r_{{\rm s}, i} \,d \log(\kappa_{{\rm s}, i}) p(\xi_i \mid \pi) p(D_{\sigma_v, i} \mid \xi_i) \biggr. \\ p(D_{{\rm LOS
}, i} \mid \xi_i) \biggl. p(\xi_i, r_{{\rm s}, i}, \log(\kappa_{{\rm s}, i}) \mid D_{\kappa, i}, D_{{\rm light}, i}) / p(\xi_i) \biggr] , 
\end{multline} 
where $p(\pi)$ is the prior on $\pi$, $p(\xi_i \mid \pi)$ is the probability of an individual system's $\xi_i$ given the parent population described by $\pi$, $p(D_{\sigma_v, i} \mid \xi_i)$ is the likelihood of the velocity dispersion measurements, $p(D_{{\rm LOS}, i} \mid \xi_i)$ is the constraint imposed by the LOS measurements, and $p(\xi_i)$ is the prior applied on $\xi_i$. 
The $p(\xi_i, r_{{\rm s}, i}, \log(\kappa_{{\rm s}, i}) \mid D_{\kappa, i}, D_{{\rm light}, i})$ term, which can also be written as $p(\tau_i, \beta_{\rm ani} \mid D_{\kappa, i}, D_{{\rm light}, i})$, represents the probability of a sampled $\xi_i$ parameter, marginalized over the individual posteriors from equation~(\ref{eq:individualposterior}).
% We then marginalize over $\tau_i$ to calculate $p(\xi_i \mid \tau_i, D_{\kappa, i}, D_{{\rm light}, i})$, which is the probability distribution of a given sampled $\xi_i$ given the system's individual posterior established in equation~(\ref{eq:individualposterior}).
% likelihood of the sampled $\gamma_{\rm in}$ to the individual posterior established by $p(\tau \mid D_{\kappa, i}, D_{{\rm light}, i})$.  We do not do the same for the $\Upsilon_{*B}$ or $\eta$ individual posteriors as we expect the kinematic information to better constrain the luminous matter-dominated inner region of the lens galaxy's mass profile \citep{newman2013}.

\begin{figure*}
    \includegraphics[width=1\linewidth]{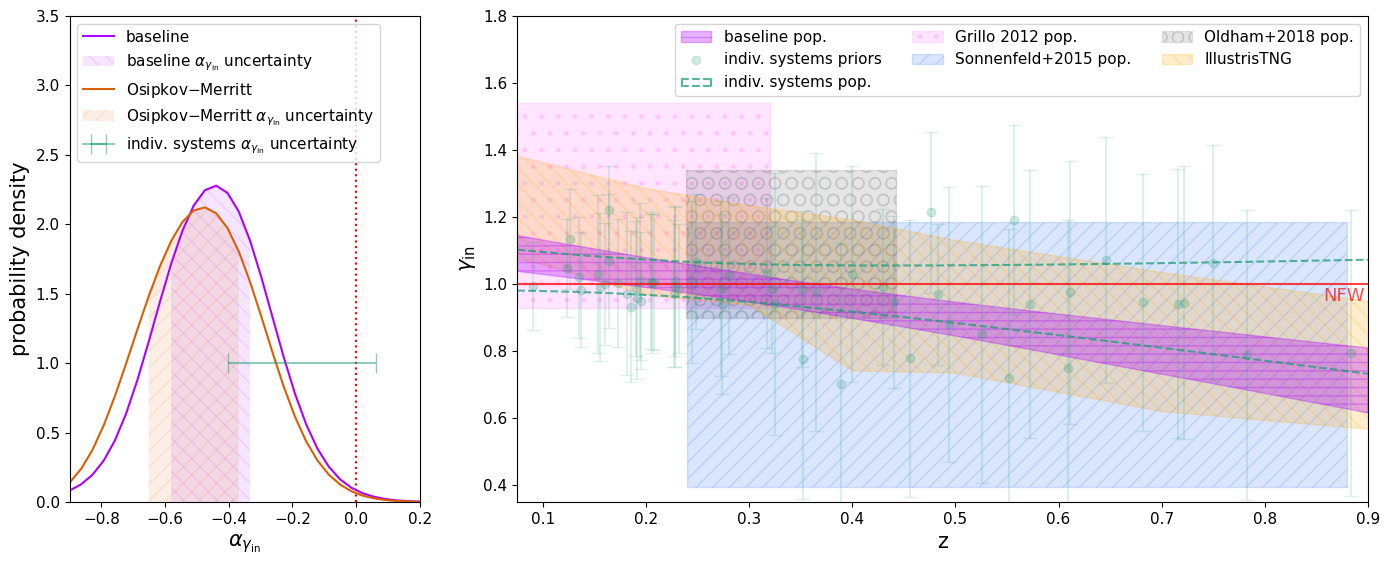}
    \caption{Left: the posterior distribution of $\alpha_{\gamma_{\rm in}}$, the population-level linear redshift evolution of the dark matter inner cusp's logarithmic slope.  We show our results assuming either a constant (purple) or Osipkov--Merritt (orange) anisotropy model, with the shaded regions corresponding to their respective uncertainties.  The red dotted line corresponds to $\alpha_{\gamma_{\rm in}}=0$ (i.e., no evolution with redshift).  Both anisotropy models are consistent with no redshift evolution of $\gamma_{\rm in}$.  
    Right: $\gamma_{\rm in}$ versus redshift, using modelled population-level parameters $\mu_{\gamma_{\rm in}}$, $\sigma_{\gamma_{\rm in}}$, and $\alpha_{\gamma_{\rm in}}$. 
    As the constant and Osipkov--Merritt anisotropy models almost completely overlap, we only plot the constant anisotropy model distribution, with the shaded purple region correspond to the 68th percentile.  
    The faint green points are the $\gamma_{\rm in}$ posteriors of each of our 56 lens sample, constrained on only lensing information, and the dashed green lines correspond to the 68th percentile population posterior from these lensing-only constraints.    
    We overlay the 68th percentile distributions from previous observational studies of elliptical galaxies \citep{grillo2012, sonnenfeld2015, O+A18}, as well as from the IllustrisTNG simulation (see Appendix~\ref{appendix:illustris}).   
    The red horizontal line corresponds to $\gamma_{\rm in}=1$ (i.e., an NFW profile). 
    \txr{Our results are in agreement with previous observational studies and with the IllustrisTNG simulation predictions.  Additionally, we find that $\gamma_{\rm in}$ is $\leq 2\sigma$ consistent with an NFW at $z \leq 0.49$, and becomes shallower than an NFW profile at higher redshifts.}}
    \label{fig:gamma_both}
\end{figure*}

%We take uniform priors on $\xi$ of $\gamma_{\rm in} \sim \mathcal{U}([0, 2.9])$ and $\log(\Upsilon_{*B}) \sim \mathcal{U}([0.1, 1.5])$.  
For the anisotropy model parameters, $\beta_{\rm ani} \sim \mathcal{U}([-0.5, 1])$ for a constant anisotropy, and $a_{\rm ani} \sim \mathcal{U}([0.2, 5.0])$ for an Osipkov--Merritt anisotropy.  We choose a constant anisotropy profile as our baseline model, as it is consistent with dynamical observables obtained for local elliptical galaxies \citep{gerhard2001, cappellari2008, cappellari2007}.  As we are sampling the $\xi$ parameters from Gaussian distributions specified by $\pi$, these priors cause $p(\xi_i \mid \pi)$ to follow a truncated Gaussian distribution (with exception to $\eta$, which is sampled from a Dirac delta function), which serve to exclude extreme values from being sampled.  As these priors are uniform and the ranges are broad enough to encompass the individual-level constraints $p(\tau \mid D_{\kappa, i}, D_{{\rm light}, i})$, we do not need to explicitly perform the division of $p(\xi_i)$ in equation~(\ref{eq:popposterior2}).  See Table~\ref{tab:pop_names} for our priors on $\pi$, the population-level parameters.  %We take priors on the population-level parameters $\pi$ of $\mu_{\gamma_{\rm in}} \sim \mathcal{U}(0.1, 2.9)$, $\sigma_{\gamma_{\rm in}} \sim \mathcal{U}(0, 2.9)$, $\gamma_{\rm in} \sim \mathcal{U}(0, 2.9)$, $\gamma_{\rm in} \sim \mathcal{U}(0, 2.9)$, and $\gamma_{\rm in} \sim \mathcal{U}(0, 2.9)$.  $\mu_{\log(\Upsilon_{*B})}$, $\sigma_{\log(\Upsilon_{*B})}$, and $\alpha_{\log(\Upsilon_{*B})}$ priors are as defined in equation~(\ref{eq:mlprior}).

%Equation~(\ref{eq:popposterior2}) should also include a $p(\xi)^{-1}$ factor, which we neglect as we do not impose any explicit prior on $\xi$ (i.e., there is a uniform prior from $-\infty$ to $\infty$ for parameters in $\xi$), since $\xi$ is already heavily regulated by the $p(\xi \mid \pi)$ and $p(\xi \mid \tau, D_{\kappa, i}, D_{L, i})$ terms.  %is the probability of a given parameterization $\xi$ from the individual distributions $\tau$ (specifically $\gamma_{\rm in}$, as we find that $\gamma_{\rm in}$ remains independent of other parameters within the relevant regime). 

\section{Results} \label{sec:results}
We test six different configuration of hierarchical hyperparameters.  Our baseline model assumes a constant anisotropy model, using the priors from Table~\ref{tab:pop_names}, $H_0 = 70$ km s$^{-1}$ Mpc$^{-1}$, and is run on our full SL2S and SLACS sample.  Furthermore, we test the same configuration but assuming an Ospikov--Merritt anisotropy model.  To test whether our results are selectively biased by our sample selection and to better compare with previous works, we run on hierarchical analysis on the SL2S and SLACS samples separately.  To check how robust our pipeline is to differing cosmology, we run two different configurations with alternative $H_0$ values of 67 and 73 km s$^{-1}$ Mpc$^{-1}$.  See Table~\ref{tab:posteriors} for the posteriors of all six of our models.  We only provide the upper 95th percentile for $\overline{\eta}$ as our perceived lower bound is likely an artifact from using lognormal distributions to represent the individual system posteriors.  Figure~\ref{fig:corner} presents the corner plot of the population-level parameter distributions of our baseline, SL2S-only, and SLACS-only models.  In Figure~\ref{fig:vd}, we present the measured velocity dispersion distributions and our resulting posteriors of the two anisotropy models, for each system.

Using the population-level posteriors, we also present the luminous and dark matter profiles for a representative, averaged system from our 56 system sample at $z_{\rm ref}$ and $\log(\sigma_{\rm v, ref})$ in Figure~\ref{fig:representative}. This average system's light profile is generated by taking the mean surface luminosity of our full sample (after being $K$-corrected, corrected for distance modulus and cosmological surface brightness dimming, and converted to absolute luminosity), at a given radius.  The $\log(\Upsilon_{*B})$ prior from equation~(\ref{eq:mlprior0}; light orange) or the population-level posterior (dark orange) is then applied to the averaged-flux profile.  For the dark matter profile, the average $R_s$ and $\log(\kappa_s)$ of the same lens sample are used. However, $\gamma_{\rm in}$ is sampled from either the population-level distribution of the individual system posteriors (which serve as priors in the hierarchical analysis; light blue) or the population-level posteriors of our hierarchical analysis (dark blue).  In Figure~\ref{fig:representative}, we qualitatively observe a tighter constraint on $\gamma_{\rm in}$ at $z_{\rm ref}$ and $\log(\sigma_{\rm v, ref})$ by applying our Bayesian hierarchical analysis; further analysis on $\gamma_{\rm in}$ and its evolution with redshift is discussed in Section~\ref{subsec:gal_evo}.  In addition, discussion on the total convergence profile and the necessity of a mass sheet is found in Section~\ref{subsec:cosmography}.
%Additionally, we see that the population $\Upsilon_{*B}$ lies above the prior.  As the prior is sampled without a preference for IMF (see Section \ref{sec:hierarchial}) this indicates that our results prefer a heavier IMF such as a Salpeter IMF, which we discuss further in Section \ref{subsec:imf}.  %Additionally, we observe a tighter constraint on $\gamma_{\rm in}$ with hierarchical analysis, which we discuss in \S\ref{sec:gamma_disc}.

%As the population-level hierarchical posteriors for the shared parameters between the constant and Osipkov--Merritt anisotropy models are similar (see Figure~\ref{fig:corner} and Table~\ref{tab:posteriors}), we only present the results assuming a constant anisotropy in the bottom panels of Figure~\ref{fig:example_decomp}, and in Figure~\ref{fig:representative}.  

%We find that $\alpha_{\log(\Upsilon_{*B})} = -0.69^{+0.07}_{-0.05}$ for an Osipkov--Merritt anisotropy, and $\alpha_{\log(\Upsilon_{*B})} = -0.67^{+0.06}_{-0.05}$ for a constant anisotropy.  Both of these posteriors agrees well with our established prior in Equation~(\ref{eq:mlprior0}).

%Overall, we do not observe a significant difference in results between an Osipkov--Merritt or constant anisotropy.  This is expected due the large aperture used for measuring velocity dispersions, and hence any differential dispersion is averaged over. 

\section{Discussion} \label{sec:discussion}
In this section, we discuss the results of our Bayesian hierarchical analysis, with an emphasis on its implications on elliptical galaxy evolution (Section~\ref{subsec:gal_evo}), IMF (Section~\ref{subsec:imf}), velocity dispersion (Section~\ref{subsec:vd_implications}), cosmography (Section~\ref{subsec:cosmography}).

\subsection{Implications on galaxy evolution}\label{subsec:gal_evo}

%\michele[I would expect us to have an update to a figure like fig.~5 of \citet{Derkenne2021} at $z\sim0.4$, showing the total slope evolution. The total slope is much more model-independent than the DM slope $\gamma_{in}$ and it would be good to show we are consistent with other results, and compare with simulations.]{
In Figure~\ref{fig:gamma_both}, we present the population distribution of $\alpha_{\gamma_{\rm in}}$ (left), and the evolution of $\gamma_{\rm in}$ over redshift (right).  From the left panel, we see that $\alpha_{\gamma_{\rm in}}$ has a $\geq 2 \sigma$ departure from no evolution $\alpha_{\gamma_{\rm in}} = 0$, and thus we find tentative evidence for evolution of the dark matter halo inner slope with redshift ($\alpha_{\gamma_{\rm in}} = -0.44^{+0.14}_{-0.15}$ for our baseline model).  \txr{Furthermore from the right panel, we find that $\gamma_{\rm in}$ is consistent with that of an NFW $\gamma_{\rm in} = 1$ (with $\mu_{\gamma_{\rm in}} = 0.97^{+0.03}_{-0.03}$ at $z_{\rm ref}$ and $\log(\sigma_{\rm v, ref})$) at redshifts $\leq 0.49$ (with $\leq 2 \sigma$ agreement), and trends towards slopes shallower than an NFW profile at higher redshifts.  
This supports the hypothesis that these dark matter overdensities conform to an NFW profile through time-accumulated baryonic processes that gradually alter the density profile over time.  }
% From the left panel, we see that $\alpha_{\gamma_{\rm in}} = 0$ is consistent within $< 1\sigma$ of our results, for all of our models.  In other words, we observe no significant linear evolution of the internal dark matter profile slope over redshift.  This is in agreement with results from the IllustrisTNG simulation (see Appendix~\ref{appendix:illustris}), which also observe no significant evolution (as shown in the right panel).  Shifting our focus onto the distribution of $\gamma_{\rm in}$ as a whole, we

We also see that our results are very consistent with previous studies of observed elliptical galaxy populations.  Our results primarily differs from previous studies as we 1.) investigate the evolution of $\gamma_{\rm in}$ with redshift, 2.) use newly observed, high $S/N$ data for the SL2S systems (which occupy higher redshifts compared to the SLACS sample), and 3.) utilize the improved and uniform models in Dinos-I.  

Our $\gamma_{\rm in}$ distribution is in agreement with the previous population studies of \citet{grillo2012}, \citet{sonnenfeld2015}, and \citet{O+A18} over their respective redshift ranges, while providing tighter, more accurate constraints across a much larger redshift range.
\txr{In addition, we also find that our results in excellent agreement with the IllustrisTNG predictions utilizing cosmological magnetohydrodynamical simulations of dark matter halo formation (see Appendix~\ref{appendix:illustris} on how we derive $\gamma_{\rm in}$ from the simulation data). }

Using our averaged light profile (see Figure~\ref{fig:representative}), we can also model for a corresponding halo response parameter ($\nu$) of our population of dark matter halos \citep{Dutton2007, shajib2021}.  Assuming that a dark matter halo initially follows an NFW profile distribution, $\nu$ is described as
\begin{equation} \label{eq:halo_response_param}
    r_f = \Gamma^{\nu} r_i,
\end{equation}
where $r_i$ and $r_f$ are the initial and final positions of a given dark matter particle, and $\Gamma$ is the contraction factor.  The halo response parameter describes the amount of adiabatic contraction ($\nu > 0$) or expansion ($\nu < 0$) of the dark matter halo due to baryonic processes.  We recovered a halo response parameter of $\overline{\nu} = 0.04^{+0.01}_{-0.01}$, and so we observe an overall contraction of the dark matter halo.  This is in slight $2.4\sigma$ tension with \citet{shajib2021}, where they recover $\mu_{\nu} = -0.06^{+0.04}_{-0.04}$ over a sample of SLACS lenses.  However, we remark that there are many differences between our analysis and that of \citet{shajib2021}.  For example, the gNFW profile used in this paper and the adiabatically contracting/expanding NFW profile are inherently different and do not directly translate between one another.  Perhaps the biggest difference between the two dark matter profiles is that the outer radius of the gNFW profile should not change considerably (see equation~\ref{gnfw_eq}), whereas the adiabatically contracting/expanding NFW profile can with varying $\nu$.  As such, the measurement of $r_f/r_i$ can systematically vary significantly outside of the scale radius between our analyses, and thus result in differing values of the halo response parameter.

Lastly, we find that our log stellar mass-to-light ratio evolves with redshift as $\alpha_{\log(\Upsilon_{*B})} = -0.57^{+0.06}_{-0.06}$ for our baseline model.  All of our models are in agreement with our prior used for $\alpha_{\log(\Upsilon_{*B})} \sim \mathcal{N}(\mu=-0.72, \sigma=0.07)$ (equation~\ref{eq:mlprior0}\txr{; with tension of at most $1.6\sigma$}), which was derived from independent fundamental plane analysis.

\subsection{Implications on IMF}\label{subsec:imf}

%Similarly to how we derive our prior on $\Upsilon_{*B}$ to not have a preference on an IMF (see Section~\ref{sec:hierarchial}), 
To infer the IMF one can compare the $\Upsilon_{*B}$ values derived from the dynamics+lensing models against the corresponding values from the stellar population $\Upsilon_{*B}^{\rm pop}$.
Here we compare our final posterior of $\log(\Upsilon_{*B})$ at $z_{\rm ref}$ and $\log(\sigma_{\rm v, ref})$ with the $\log(\Upsilon_{*B}^{\rm pop})$ distribution of elliptical galaxy lenses measured by \citet{auger2009} converted to the same redshift \citep[assuming a given IMF, and accounting for the redshift dependence from][]{treu2005}.  

\begin{figure}
    \includegraphics[width=1\linewidth]{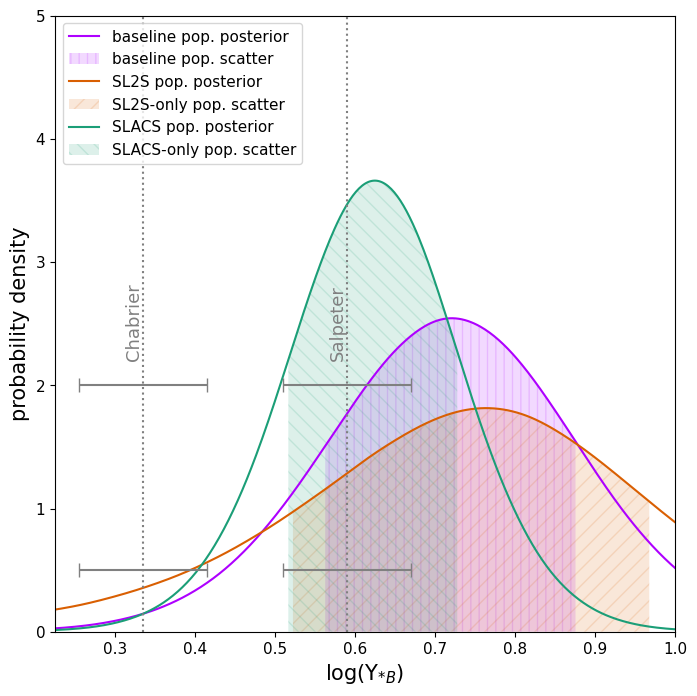}
    \caption{The posterior distribution of $\log(\Upsilon_{*B})$ at $z_{\rm ref}$ and $\log(\sigma_{\rm v, ref})$, sampled from the population-level distributions of $\mu_{\log(\Upsilon_{*B})}$ and $\sigma_{\log(\Upsilon_{*B})}$.  We show our results of our hierarchical analysis applied on our full sample (purple), only the SL2S systems (orange), or only the SLACS systems (green).  The shaded region illustrates the 68th percentile of the $\log(\Upsilon_{*B})$ posterior, for their respective distribution.  Also shown is the 68th percentile uncertainty of our $\mu_{\log(\Upsilon_{*B})}$ prior (see equation~\ref{eq:mlprior0}), adjusted by $\pm \log(1.8)/2$ for a Salpeter or Chabrier IMF, respectively.  All models strongly prefer a Salpeter IMF over a Chabrier IMF.}
    \label{fig:ml_hist}
\end{figure}

We present this comparison in Figure~\ref{fig:ml_hist}, for our baseline, SL2S-only, and SLACS-only models.  All models favour a heavier, Salpeter-like IMF.  This is in agreement with similar analyses on elliptical galaxy lenses \citep[e.g.,][]{Treu10,sonnenfeld2012, shajib2021}, with results from galaxy dynamics, which indicate a Salpeter-like IMF for galaxies with large velocity dispersion \citep[e.g.,][]{Cappellari2012, Cappellari2013, Li2017, Lu2024} and from analysis of the spectra features \cite[e.g.,][]{Dokkum2010, vandokkum12,conroy2012, Spiniello2012, Smith20}, which indicates that elliptical galaxies favour an IMF heavier (Salpeter-like) than what is observed of the Milky Way \citep[Chabrier;][]{chabrier, chabrier2023}.  

As theoretical studies have shown, a bottom-heavy IMF can be indicative of higher fragmentation rates of molecular clouds, and thus a larger population of dwarf stars \citep{Hopkins2013, Chabrier2014}. We note that the observed stellar mass to light gradient could arise not only from trends in radial stellar population age and metallicity, but also potentially in the IMF as suggested from stellar population analysis \citep[e.g.,][]{MartinNavarro2015, vandokkum17, Parikh2018, LaBarbera2019}.  Our data are not sufficient to distinguish the two hypotheses. However, recent analysis for the MaNGA galaxy survey found no dynamical evidence of the IMF gradients indicated by the studies of spectral absorptions \citep{Lu2024}. This suggests either systematic issues in either technique or that the IMF gradients indicated by the stellar population do not affect the $\log(\Upsilon_{*B}^{\rm pop})$ as predicted by simple assumptions on the IMF shape.

\subsection{Implications on velocity dispersion measurements and dependence}\label{subsec:vd_implications}

We find that the linear relationship between $\gamma_{\rm in}$ and $\log(\sigma_{\rm v})$ is $\epsilon_{\gamma_{\rm in}} = 0.39^{+0.39}_{-0.41}$ for our baseline model.  In other words, our result is consistent with no linear trend between the inner dark matter logarithmic slope and the logarithmic velocity dispersion; this result holds true across all of our models.  We caution, however, that our range in velocity dispersion is relatively small, and thus the errors are large. Studies covering a larger range of stellar velocity dispersion may recover a significant trend.

As for the the linear relation between $\log(\Upsilon_{*B})$ and $\log(\sigma_{\rm v})$, we find $\epsilon_{\log(\Upsilon_{*B})} = 0.56^{+0.28}_{-0.28}$ for our baseline model, and $\epsilon_{\log(\Upsilon_{*B})} = 1.06^{+0.46}_{-0.39}$ for our SLACS-only model.  A similar quantity is measured in \citet{treu2010b} and \citet{Sonnenfeld2017}, but instead of finding the correlation between $\log(\Upsilon_{*B})$ and the logarithmic velocity dispersion, they probe the correlation between $\log(M^{\rm LD}_{\rm *, Ein}/M^{\rm SPS}_{\rm *, Ein})$ and $\log(\sigma_{\rm v})$.  $M^{\rm LD}_{\rm *, Ein}$ is the stellar mass inferred by lensing and dynamical models, and $M^{\rm SPS}_{\rm *, Ein}$ is the stellar mass inferred from stellar populations synthesis models within the Einstein radius; the ratio of which is referred to as the ``IMF mismatch'' parameter.  \citet{treu2010b} measures a slope between the logarithmic IMF mismatch parameter and $\log(\sigma_{\rm v})$ of $1.31 \pm 0.16$ in a SLACS sample, while \citet{Sonnenfeld2017} measures $0.81 \pm 0.21$ in a SL2S and SLACS sample.  \txr{While this parameterization is not perfectly analogous to the $\epsilon_{\log(\Upsilon_{*B})}$ used in this paper (e.g., the ``IMF mismatch'' parameter used in \citet{treu2010b} and \citet{Sonnenfeld2017} does not account for a mass-to-light gradient), both of these measurements are nonetheless $\leq 1 \sigma$ consistent with our results for using corresponding dataset.}

\subsection{Implications on time-delay cosmography}\label{subsec:cosmography}

In our models with varying $H_0$ (67, 70, and 73 km s$^{-1}$ Mpc$^{-1}$; see Table~\ref{tab:posteriors}), we find that our results are self-consistent and robust against an assumed cosmology (specifically, with $H_0$).  This is expected, as our analysis should only depend on the ratio of cosmological distances.  \txr{Additionally, as our mass-concentration relation assumption is only significantly affected by $\sigma_8$ and $w_0$ \citep[e.g.,][]{lopez-cano2022}, this also means that our results can be used independently of $H_0$}, and can be applied to lensed cosmography discussions without systematic bias.  Of course, this also means that our implications on galaxy evolution (Section~\ref{subsec:gal_evo}), IMF (Section~\ref{subsec:imf}), and velocity dispersion (Section~\ref{subsec:vd_implications}) remain robust against our choice of $H_0$.

From Table~\ref{tab:posteriors}, we find our $\overline{\eta}$ posterior to be $\overline{\eta}  \leq 5 \times 10^{-5}$ for our baseline model, and $\overline{\eta} \leq 1.3 \times 10^{-4}$ for our SLACS-only model.  In comparison to \citet{shajib2021} which finds that $\exp(\mu_{\log(\eta)}) \leq 1.7\times 10^{-2}$ on an SLACS lens sample, our results are not only in agreement with \citet{shajib2021}, but also provides tighter upper bounds on the stellar mass-to-light ratio gradient \txr{and stronger evidence for a negligible stellar mass-to-light ratio gradient at the population level}.  We also find that our stellar orbits are generally isotropic, with no evidence of a departure from isotropy.  

% This is in agreement with what \citet{shajib2021} measures in their analysis of SLACS systems ($\exp(\mu_{\log(\eta)}) \leq 1.7\times 10^{-2}$; which also agrees well with our $\overline{\eta} = \leq 8.1 \times 10^{-3}$ measurement of our SLACS-only model).  However, we nevertheless find a small, but non-negligible, average stellar mass-to-light gradient.  
\txr{From Figure~\ref{fig:gamma_both}, while our $\gamma_{\rm in}$ distribution at $z \leq 0.49$ is consistent with a NFW profile at the $2 \sigma$ level, results indicate an significantly shallower inner slope at higher redshifts.  At the $3 \sigma$ level, the redshift range where $\gamma_{\rm in} \approx 1$  expands to $z \leq 0.83$.}  As the lensed quasars used by the TDCOSMO collaboration are at centred at approximately $z_{\rm l} \sim 0.5$, this sample can be seen as broadly consistent (within 1 to \txr{3}$\sigma$) with using an NFW profile.  
% However, we find that accounting for the two extra degrees of freedom is ideal when modelling using precision hybrid models: a stellar mass-to-light gradient $\eta$ for the luminous matter profile, and an independent inner slope $\gamma_{\rm in}$ for the dark matter profile of elliptical galaxy lenses (especially for those at lower redshifts).

We find that the power-law profile remains a robust method of modelling the total surface density profile for our sample of galaxies.  From our averaged profile, for a generous radius range of expected lensed image locations ($0.25 R_{\rm E}$ to $4 R_{\rm E}$), our total mass posterior profile is in excellent agreement with a power-law (Figure~\ref{fig:representative}, green shaded region and purple line).  

To quantify this agreement and to probe the mass sheet transformation (MST), we parameterize a convergence model using a power-law with a mass sheet as shown by \citet{Falco1985}:
\begin{equation}
    \kappa_{\lambda} = \lambda \kappa_{\rm PL} + (1-\lambda),
\end{equation}
where $\kappa_{\rm PL}$ is the power-law convergence profile (as a function of the Einstein radius and $\gamma$) and $\lambda$ is the internal MST parameter (where $\lambda = 1$ is equivalent to having no mass sheet).  We fit this convergence profile to the $\mu$ total matter convergence profile (shown as the dark shaded green region in Figure~\ref{fig:representative}) within the range of $0.25 R_{\rm E}$ to $4 R_{\rm E}$, by minimizing the average variance marginalized over log-space.  By doing we, we measure the mean internal MST parameter of our sample to be $\overline{\lambda} = 0.96 \pm 0.03$, and a mean total density logarithmic slope of $\overline{\gamma} = 2.24 \pm 0.14$.  We find that a power-law profile can well describe the total matter convergence within this broad lensing regime.  Our $\overline{\lambda}$ result is consistent with those found by Dinos-I when assuming a constant anisotropy model ($\mu_{\lambda} = 0.91^{+0.10}_{-0.09}$), and shows that our population of elliptical galaxy lenses are consistent (\txr{$\leq 1.2 \sigma$}) with having no mass sheet present.  Additionally, our $\overline{\gamma}$ agrees well \txr{(at or within $1 
\sigma$)} with other average logarithmic slope measurements from dynamical models, such as $\mu_{\gamma} = 2.19 \pm 0.03$ from \citet{Cappellari2015} and $\mu_{\gamma} = 2.078 \pm 0.027$ from \citet{auger2010}.  

\txr{We note that these results, especially that of our $\overline{\lambda}$ posterior, are made under the assumption that the total matter profile of the elliptical galaxy is composed of a gNFW dark matter profile and a luminous surface density profile according to equation~(\ref{eq:lum_kappa}).  In contrast, Dinos-I estimates a population posterior of $\mu_{\lambda} = 0.91^{+0.10}_{-0.09}$, while explicitly allowing for maximal flexibility of the total matter profile and more model-robust result.  Nevertheless, our model-specific result is in good agreement with the results of Dinos-I.}

As the MST is an ongoing obstacle in accurately determining $H_0$ measurements through the study of lensed quasars, this result indicates that the systematic bias imposed by a MST may be minimal (when using a power-law profile to model a lens convergence).  As the measured Hubble constant is linearly related to $\lambda$, tighter constraints on cosmology can be achieved when we reduce the theoretical effects of the MST.

% \begin{multline}
%     \overline{\sigma^2} \equiv \frac{1}{2}  \int_{-1}^{1} dx \, \left( \frac{\kappa_{\lambda}(4^x R_{\rm E}) - \mu_{\kappa_{\rm T}}(4^x R_{\rm E})}{\sigma_{\kappa_{\rm T}}(4^x R_{\rm E})} \right)^2 \\
%     = \frac{1}{4\ln 2} \int_{0.25 R_{\rm E}}^{4 R_{\rm E}} \frac{dr}{r} \left( \frac{\kappa_{\rm PL}(r) - \mu_{\kappa_{\rm T}}(r)}{\sigma_{\kappa_{\rm T}}(r)} \right)^2,
% \end{multline}
% where $\kappa_{\rm PL}(r)$ is the power-law convergence profile, $\mu_{\kappa_{\rm T}}(r)$ is mean of the total mass convergence profile, and $\sigma_{\kappa_{\rm T}}(r)$ is scatter of the total mass convergence profile (accounting for uncertainty of $\mu$ and intrinsic scatter $\sigma$ for all population-level parameters).  As $\overline{\sigma^2} << 1$, we find that a power-law profile can well describe the total matter convergence within the relevant lensing regime of our population of elliptical galaxy lenses.

\section{Conclusion} \label{sec:conclusion}
% In this work, we model 21 SL2S lenses using recent \textit{HST} WFC3 F475X observations, which span redshifts from $z_{\rm l}=0.325$ to $0.884$.  As the F475X filter is bluer compared to other optical bands, the features of the source arcs are accentuated while the lens galaxy flux is diminished.  This allows us to obtain tighter statistical and systematic constraints on $R_E$ and $\gamma$ than previous studies.  

% By combining these systems with four additional SL2S and 33 SLACS lensing systems from DINOS~I, we can further augment our redshift range from $z_{\rm l}=0.090$ to $0.884$.  This large range in redshift allows us to measure population-level parameter trends (with relation to $z$) and distributions, which we achieve through Bayesian hierarchical analysis.  

% We parameterize the stellar luminous matter profile with a stellar mass-to-light ratio $\Upsilon_{*B}$ and gradient $\eta$, while we model the dark matter profile as a general NFW, %\michele[As commented before, We should discuss the more robust total slopes before jumping to the more model-dependent DM slopes]{
% which incorporates an parameterizable inner slope $\gamma_{\rm in}$.  With this model, we constrain over the lens light profiles, model-independent strong lensing constraints, velocity dispersion, and LOS measurements for each system, as well as dark matter $M$--$c$ relation and mass-to-light ratio priors.  

In this work, we study the population statistics of dark and luminous matter profiles from a sample of 56 SL2S and SLACS elliptical galaxy lenses, spanning a redshift of $0.090 \leq z \leq 0.884$.  The parameters include the inner logarithmic slope of the dark matter surface density profile ($\gamma_{\rm in}$), the stellar mass-to-light ratio ($\Upsilon_{*, B}$), the stellar mass-to-light gradient ($\eta$), and anisotropic parameters.  Due to our large and diverse sample, we can also accurately probe the correlation of $\gamma_{\rm in}$ and $\log(\Upsilon_{*, B})$ with both $z$ and $\log(\sigma_{\rm v})$.  From our analysis, we better our understanding of elliptical galaxy populations, from their properties to their origin and evolution of their surface density profiles, as well as their implications for cosmology.  We present the main results of our paper as the following:
\begin{itemize}
    \item We model 21 lenses (nine of which previously unobserved by \textit{HST}), with new high resolutions observations in the F475X.  As the imaging $S/N$ is significantly higher than those used in previous models, our measurements of the lensing parameters (such as $R_{\rm E}$ and $\gamma$) are more precise and accurate for these 21 systems.
    \item We place tighter constraints on $\mu_{\gamma_{\rm in}}=0.97^{+0.03}_{-0.03}$ with $\leq 0.07$ intrinsic scatter (at $z=0.353$, $\log(\sigma_{\rm v})=2.42$, and assuming a constant anisotropy model), especially at higher redshifts as seen in Figure~\ref{fig:gamma_both}.  These results are consistent (within $1 \sigma$) with previous studies on elliptical galaxy \txr{populations, as well as with the IllustrisTNG simulation predictions.}
    \item Our results measure a $\gamma_{\rm in}$ redshift evolution of $ d \gamma_{\rm in} / d z \equiv \alpha_{\gamma_{\rm in}} = -0.44^{+0.14}_{-0.15}$ for our baseline model.  Combined with the result of the previous bullet point, the overall trend of our result is that our population dark matter distribution \txr{is $\leq 2 \sigma$ consistent with an NFW profile slope ($\gamma_{\rm in} = 1$) at lower redshifts of $0.09 \leq z \leq 0.49$ (with the range being $0.09 \leq z \leq 0.83$ for $\leq 3 \sigma$ consistency), while being shallower than that of an NFW profile at higher redshifts up to $z=0.9$.  Therefore, we find that the TDCOSMO collaboration lensed quasars (with $z_{\rm l} \sim 0.5$) are broadly consistent with an NFW dark matter profile (i.e., within 1 to 3$\sigma$).}
    \item We determine that $\gamma_{\rm in}$ and $\log(\sigma_{\rm v})$ have a correlation of $\epsilon_{\gamma_{\rm in}} = 0.39^{+0.39}_{-0.41}$ using our baseline model\txr{, which is a new result which previously has not been studied.}
    \item We find the correlation between $\log(\Upsilon_{*B})$ and $\log(\sigma_{\rm v})$ to be $\epsilon_{\log(\Upsilon_{*B})} = 0.56^{+0.28}_{-0.28}$ for our baseline model. \txr{This is in agreement with previous works using fundamental plane analysis}.  %Additionally, we measure an SDSS $\sigma_{\rm v}$ fractional systematic uncertainty of $\zeta_{{\rm SDSS}, {\rm sys}} = 0.13^{+0.02}_{-0.02}$.
    \item Our population of massive elliptical lenses strongly favour a Salpeter IMF over a Chabrier IMF (see Figure~\ref{fig:ml_hist}), which agrees well with previous analyses over similar elliptical galaxy populations.
    % \item Our results are robust to different anisotropy models, as verified by a baseline radially-constant anisotropy model and an Osipkov--Merritt anisotropy model (see Table~\ref{tab:posteriors}).
    \item Our results are robust against assumed values of $H_0$ (as our hierarchical analysis only relies on the ratio of cosmological distances) and different anisotropy models.  We verify this by checking the self-consistency of our results with $H_0 =$ 67, 70, and 73 km s$^{-1}$ Mpc$^{-1}$, as well as with a radially-constant anisotropy model and an Osipkov--Merritt anisotropy model (see Table~\ref{tab:posteriors}).
    \item Within $0.1 < z < 0.9$, we find a small\txr{, almost-negligible distribution of $\overline{\eta} \leq 5 \times 10^{-5}$}, providing tighter upper bounds to the stellar mass-to-light gradient within elliptical galaxy populations.  
    \item Our average total mass convergence is in agreement with that of a power-law profile within a generous range of $0.25 R_{\rm E} - 4 R_{\rm E}$ (see Figure~\ref{fig:representative}).  We find our mean internal MST parameter to be consistent with no mass sheet (within $1.2 \sigma$; $\overline{\lambda} = 0.96 \pm 0.03$), and our mean total density logarithmic slope to be $\overline{\gamma} = 2.24 \pm 0.14$.
    % \item \txr{averaged total gamma}
\end{itemize}

% Our results places tighter constraints on $\gamma_{\rm in}$, especially at higher redshifts, while being consistent (within $1 \sigma$) with previous studies on elliptical galaxy samples.  However, we observe a shallower slope of $2 \sigma$ compared to IllustrisTNG simulation, and a steeper slope of $1-2 \sigma$ compared to the conventional NFW profile.  Our results find that the redshift evolution of $\gamma_{\rm in}$ is consistent with no evolution.
% Additionally, our population of massive elliptical lenses strongly favour a Salpeter IMF over Chabrier, which agrees well with previous analyses over similar elliptical galaxy populations.  
% Within $0.1 \le z \le 0.9$, we find a non-negligible distribution of $\overline{\eta}$, indicating that a stellar mass-to-light gradient is necessary to model the luminous matter profile for our population of lenses.  Finally, our total mass convergence seem in strong agreement with that of a power-law profile within a generous range of $0.25 R_{\rm E} - 4 R_{\rm E}$.

In the near future, upcoming surveys such as the Vera C. Rubin Observatory Legacy Survey of Space and Time (LSST), the \textit{Nancy Grace Roman Space Telescope}, and the \textit{Euclid}-wide surveys are expected to boost the current number of strong lenses by two to three orders of magnitudes \citep{verma2019, Oguri2010, collett2015}.  %Also, with the \textit{JWST} to observe many of these lenses at a significantly higher resolution, it becomes possible to probe a much more extensive sample of massive elliptical lenses at deeper redshifts and with higher precision.  
With this, tighter constraints on the redshift evolution of elliptical galaxies, and an even-better understanding of the formation of dark matter halos can be achieved.

\section*{Acknowledgements}
The authors thank Dhayaa Anbajagane for helpful discussions about the IllustrisTNG simulation. This research is based on observations made with the NASA/ESA \textit{HST} obtained from the Space Telescope Science Institute, which is operated by the Association of Universities for Research in Astronomy, Inc., under NASA contract NAS 5–26555. These observations are associated with program HST-GO-17130.
Support for this program was provided by NASA through a grant from the Space Telescope Science Institute, which is operated by the Association of Universities for Research in Astronomy, Inc., under NASA contract NAS 5-03127.
%AJS
This work was also supported by NASA through the NASA Hubble Fellowship grant HST-HF2-51492 awarded to AJS by the Space Telescope Science Institute, which is operated by the Association of Universities for Research in Astronomy, Inc., for NASA, under contract NAS5-26555.

This research made use of \textsc{lenstronomy} \citep{birrer2018}, \textsc{Numpy} \citep{harris2020}, \textsc{scipy} \citep{virtanen2020}, \textsc{astropy} \citep{astropy2013, astropy2018, astropy2022}, \textsc{jupyter} \citep{kluyver2016}, \textsc{matplotlib} \citep{hunter2007}, \textsc{sextractor} \citep{bertin1996}, \textsc{emcee} \citep{emcee}, \textsc{corner} \citep{corner}, \textsc{kcorrect} \citep{blanton2007}, and \textsc{hierarc} \citep{birrer2020}.

%%%%%%%%%%%%%%%%%%%%%%%%%%%%%%%%%%%%%%%%%%%%%%%%%%
\section*{Data availability}
The \textit{HST} data described here
may be obtained from the Mikulski Archive for Space Telescopes (MAST) archive at \url{https://dx.doi.org/10.17909/8cfx-sq72}.  The Canada-France-Hawaii Telescope Legacy Survey final data release is available here \url{https://www.cfht.hawaii.edu/Science/CFHTLS/}.  Preprocessing, modelling, and hierarchical analysis notebooks used in this work are presented on the GitHub page \url{https://github.com/williyamshoe/dinos2}.  The lens modelling code Lenstronomy (\url{https://github.com/lenstronomy/lenstronomy}) and the hierarchical analysis code HierArc (\url{https://github.com/sibirrer/hierArc}) are also publicly available on GitHub.  



\bibliographystyle{mnras}
\bibliography{example} % if your bibtex file is called example.bib

%\begin{thebibliography}{99}
%\end{thebibliography}

%%%%%%%%%%%%%%%%%%%%%%%%%%%%%%%%%%%%%%%%%%%%%%%%%%

%%%%%%%%%%%%%%%%% APPENDICES %%%%%%%%%%%%%%%%%%%%%

\appendix
\section{SL2S lensing and light models} \label{appendix:models}

\begin{figure*}
 \includegraphics[width=7 in]{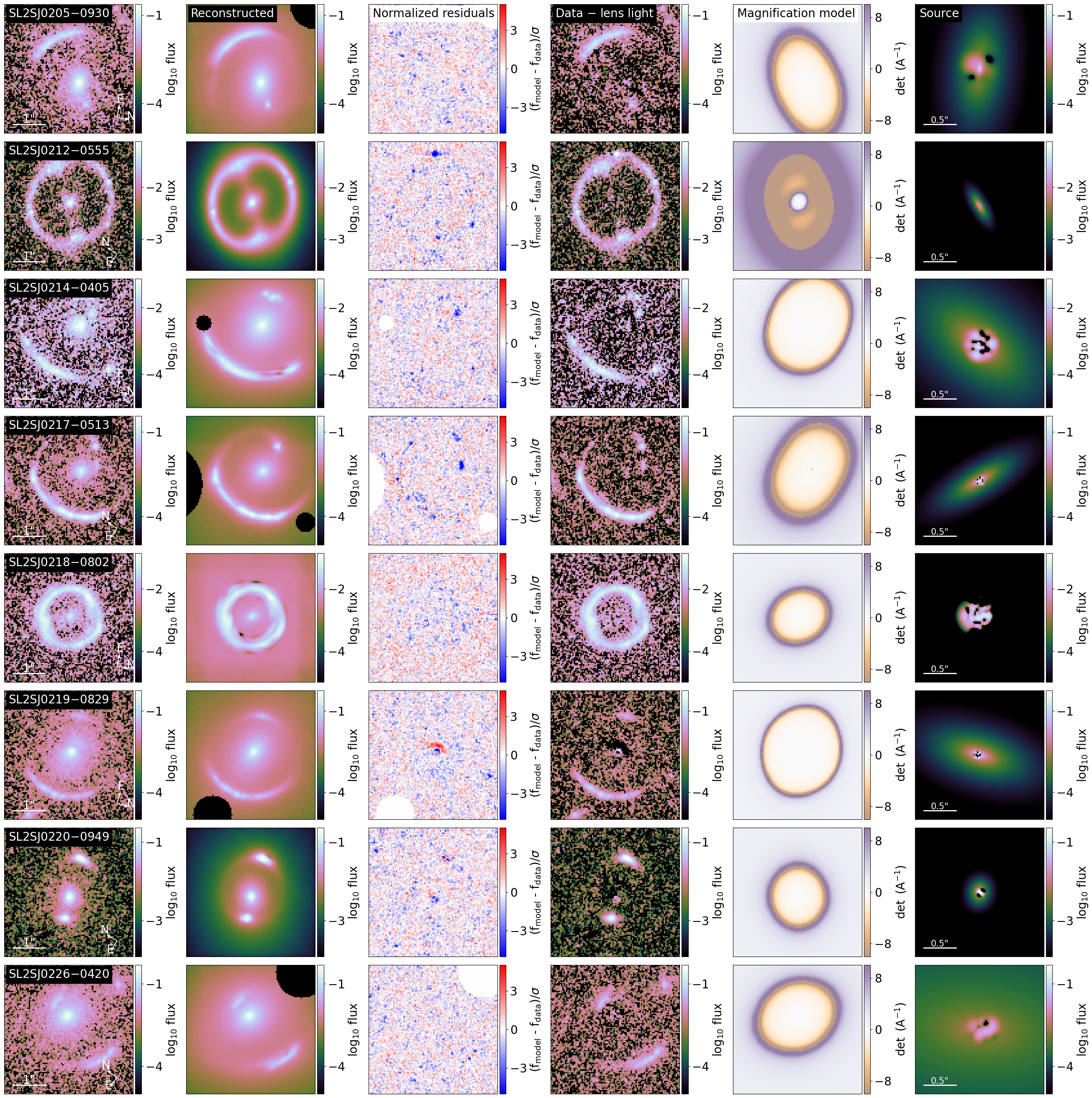}
 \caption{Lens models for the first eight out of the 21 SL2S lenses in our sample, using the \textit{HST} F475X imaging.  The first column shows the \textit{HST} F475X observed image.  The second column shows the reconstructed image from our best-fitting model.  The third column shows the normalized residuals that are minimized in fitting for the model.  The fourth column shows the \textit{HST} observed image subtracted by the best fitting model's lens light profile, illustrating the lensed arcs.  The fifth column shows the magnification map of the lensing profile of the best-fitting model.  The sixth column shows the reconstructed source of the best-fitting model.  The remaining systems are shown in Figures~\ref{fig:sing1} and \ref{fig:sing2}.}
 \label{fig:sing0}
\end{figure*}

\begin{figure*}
 \includegraphics[width=7 in]{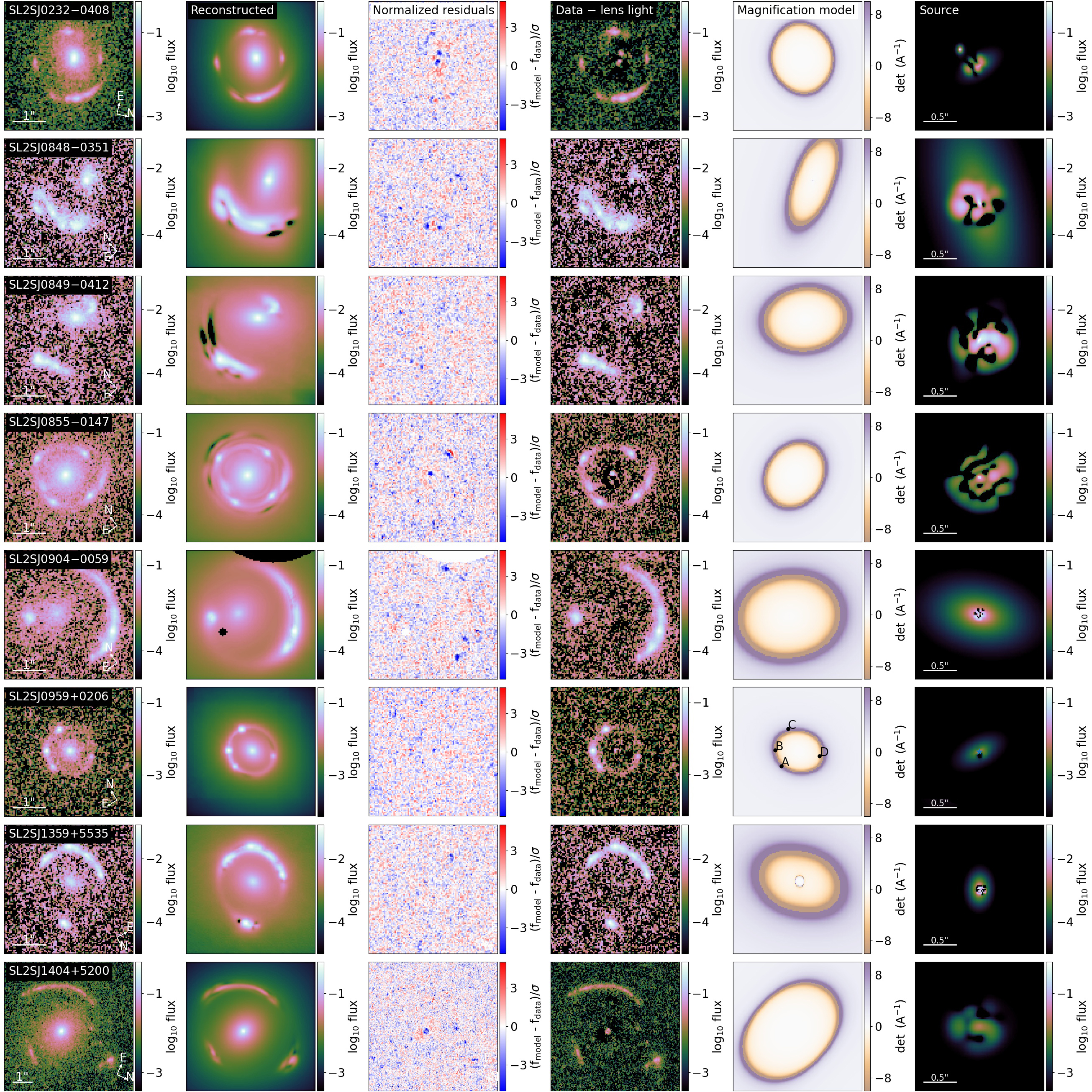}
 \caption{Lens models for the next eight (of 21) SL2S lenses in our sample, using the \textit{HST} F475X observations.  See the caption of Figure~\ref{fig:sing0} for the full description.}
 \label{fig:sing1}
\end{figure*}

\begin{figure*}
 \includegraphics[width=7 in]{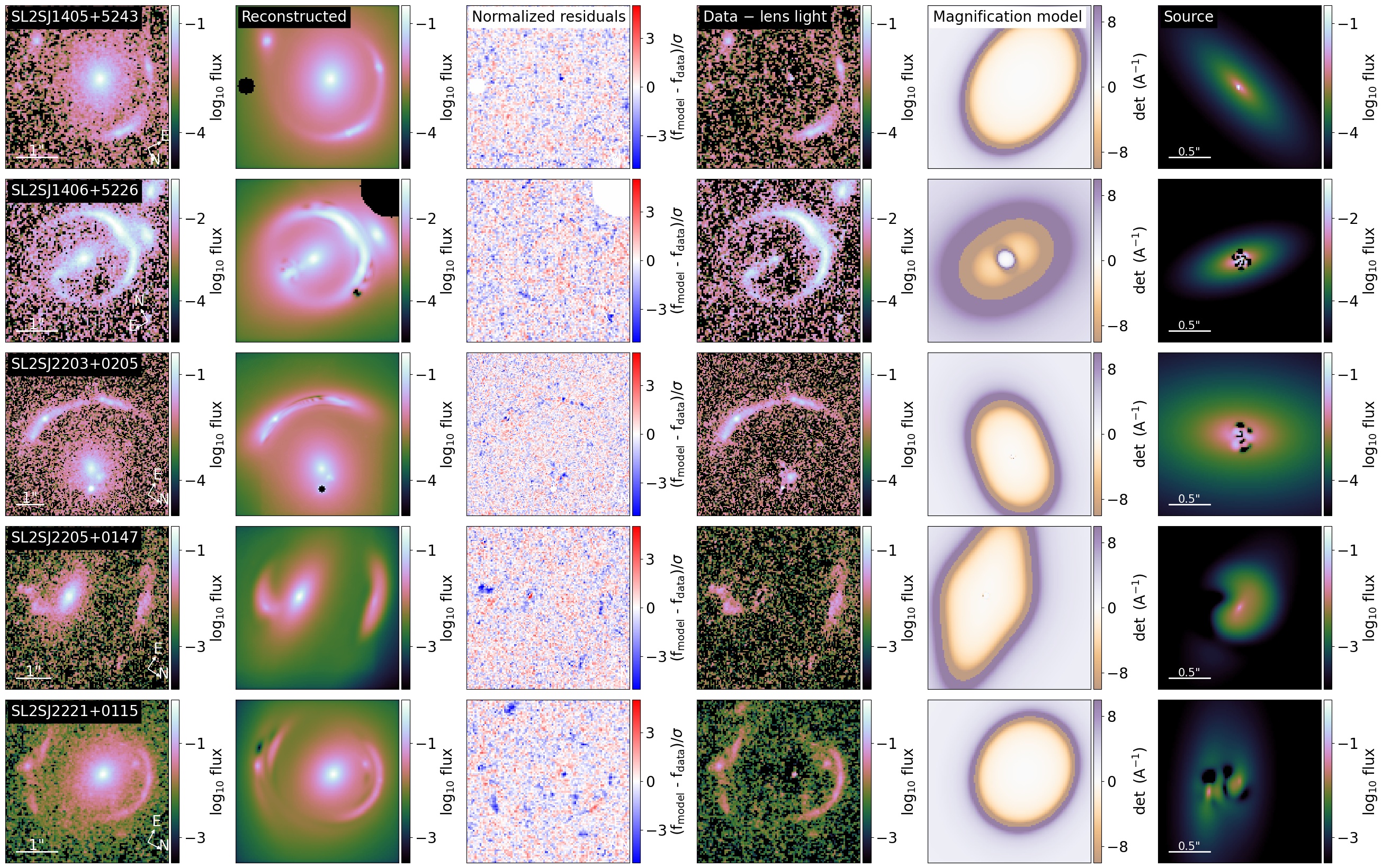}
 \caption{Lens models for the last five (of 21) SL2S lenses in our sample, using the \textit{HST} F475X observations.  See the caption of Figure~\ref{fig:sing0} for the full description.}
 \label{fig:sing2}
\end{figure*}

\begin{figure*}
 \includegraphics[width=5 in]{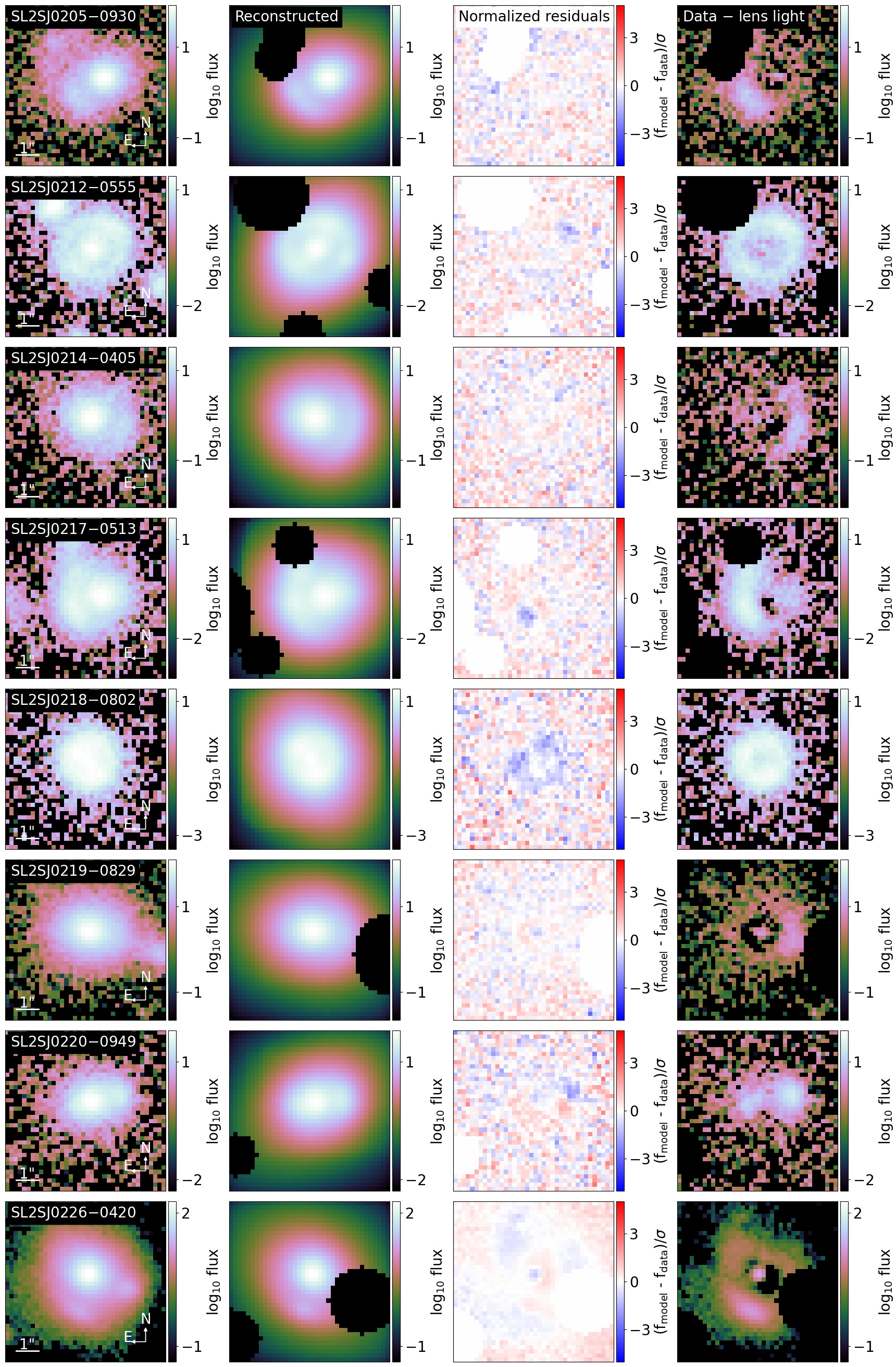}
 \caption{Lens light models for the first eight out of the 21 SL2S lenses in our sample, using the CFHT \ipr-band observations.  The first column shows the CFHT \ipr-band observed image.  The second column shows the reconstructed image from our best-fitting model.  The third column shows the normalized residuals that are minimized in fitting for the model.  The fourth column shows the \textit{HST} observed image subtracted by the best fitting model's lens light profile, illustrating the lensed features.  As we are primarily interested in the CFHT \ipr-band for lens galaxy light profiles, the lensing parameters of these models are heavily restricted by the \textit{HST} models and the source galaxy light is modelled with a single Sérsic profile.  Therefore, we do not show the magnification map or the reconstructed source.  The remaining systems are shown in Figures~\ref{fig:mult1} and \ref{fig:mult2}.}
 \label{fig:mult0}
\end{figure*}

\begin{figure*}
 \includegraphics[width=5 in]{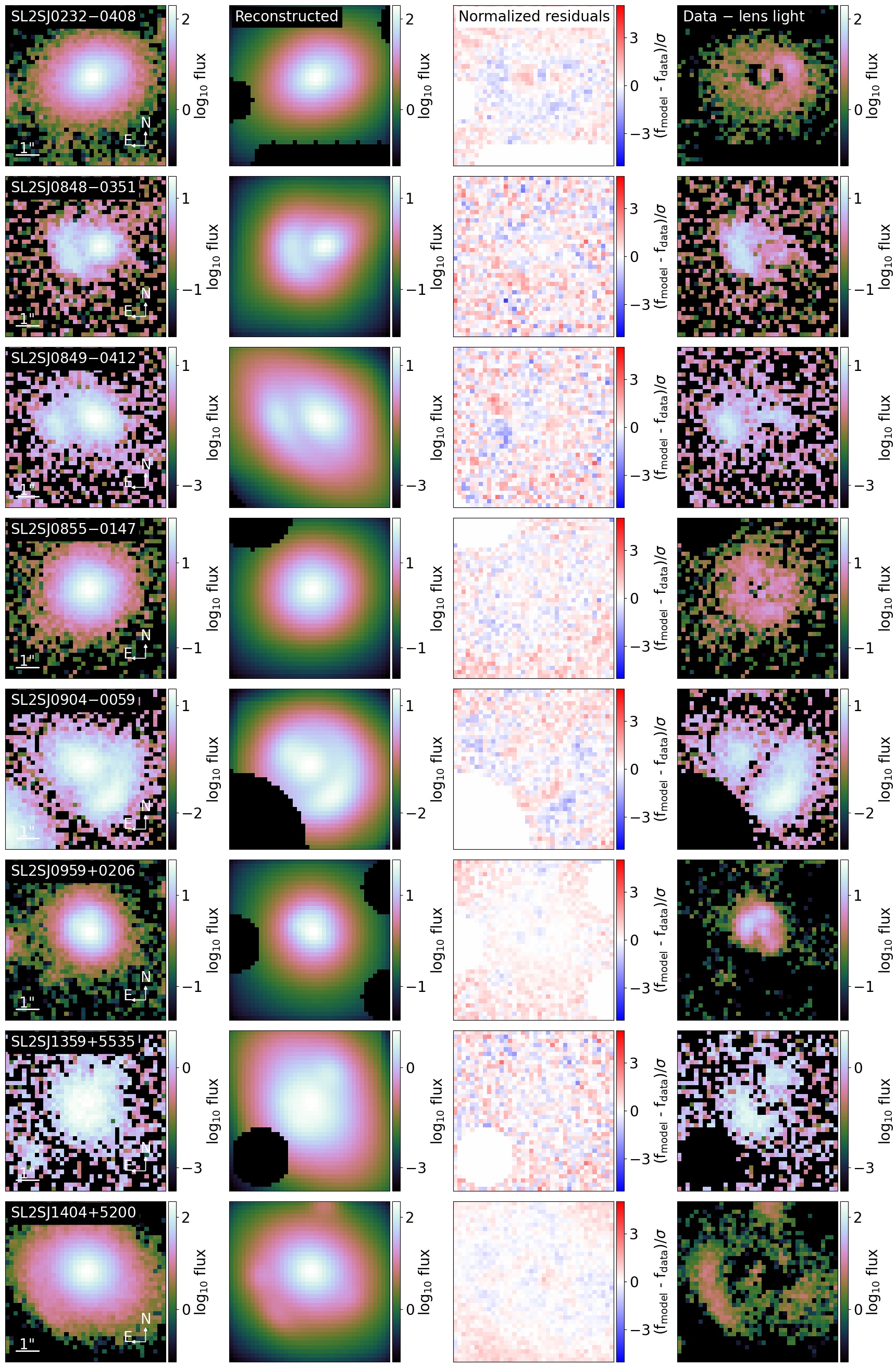}
 \caption{Lens light models for the next eight (of 21) SL2S lenses in our sample, using the CFHT \ipr-band observations.  See the caption of Figure~\ref{fig:mult0} for the full description.}
 \label{fig:mult1}
\end{figure*}

\begin{figure*}
 \includegraphics[width=5 in]{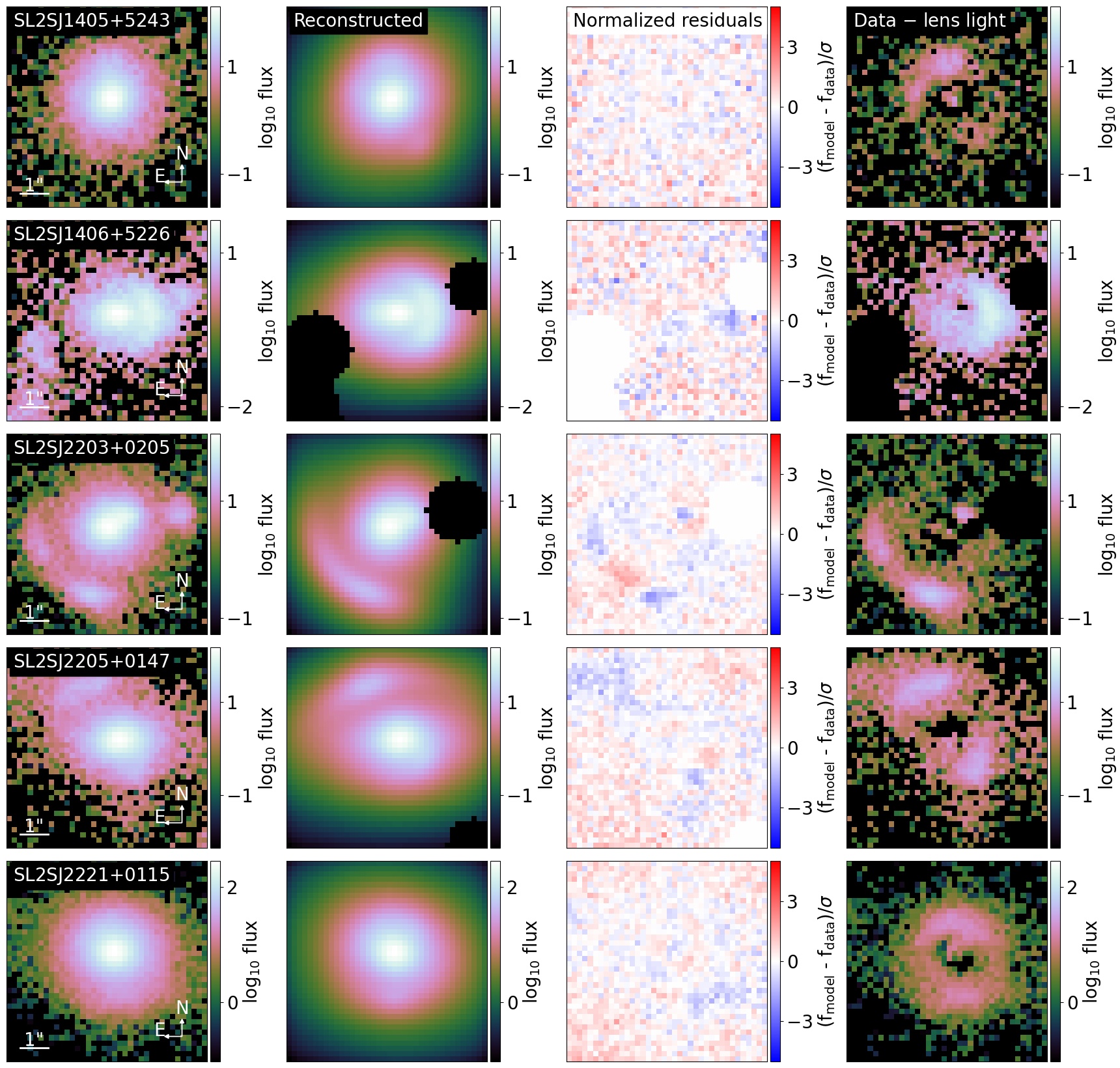}
 \caption{Lens light models for the last five (of 21) SL2S lenses in our sample, using the CFHT \ipr-band observations.  See the caption of Figure~\ref{fig:mult0} for the full description.}
 \label{fig:mult2}
\end{figure*}

Here, we show the \textit{HST} lensing (Figures~\ref{fig:sing0}, \ref{fig:sing1}, and \ref{fig:sing2}) and CFHT \ipr-band light (Figures~\ref{fig:mult0}, \ref{fig:mult1}, and \ref{fig:mult2}) models for the 21 SL2S systems.  For a majority of systems, we fit for the model scheme outlined in Section~\ref{sec:modelling}.  The following systems do not fully conform to this modelling scheme:

\paragraph*{SL2SJ0232--0408} We find it necessary to model for a second source galaxy.  Thus, we add a second Sérsic light profile to the source light profile.

\paragraph*{SL2SJ0959+0206} \txr{For this system, we changed the source light model from a S\'ersic + shapelets basis to S\'ersic + point source + shapelets basis.  This change is in response to how dense the source galaxy core is, which we speculate could introduce fluctuation in the point-source-like images.  To account for this, we add additional degrees of freedom into our model to account for these potential flux anomalies.  These point sources locations are labelled in the magnification map and source plane image in Figure~\ref{fig:sing1}.}

\paragraph*{SL2SJ2203+0205} For this system, there is an interloping point source within one arcsecond of the lens galaxy.  We model for this additional light profile using a point source light profile.

\paragraph*{SL2SJ2221+0115}  Like with SL2SJ0232--0408, we find it necessary to model for a second source galaxy, and so a second Sérsic light profile is added to the source light profile.

\section{Luminous and Dark Matter Posteriors} \label{appendix:decomps}

We present the luminous and dark matter decomposition distributions for the 56 SL2S and SLACS strong lens sample (Figures~\ref{fig:decomp1}, \ref{fig:decomp1}, \ref{fig:decomp2} and \ref{fig:decomp3}) used in our hierarchical analysis, as discussed in Section~\ref{sec:hierarchial}.

\begin{figure*}
 \includegraphics[width=1\linewidth]{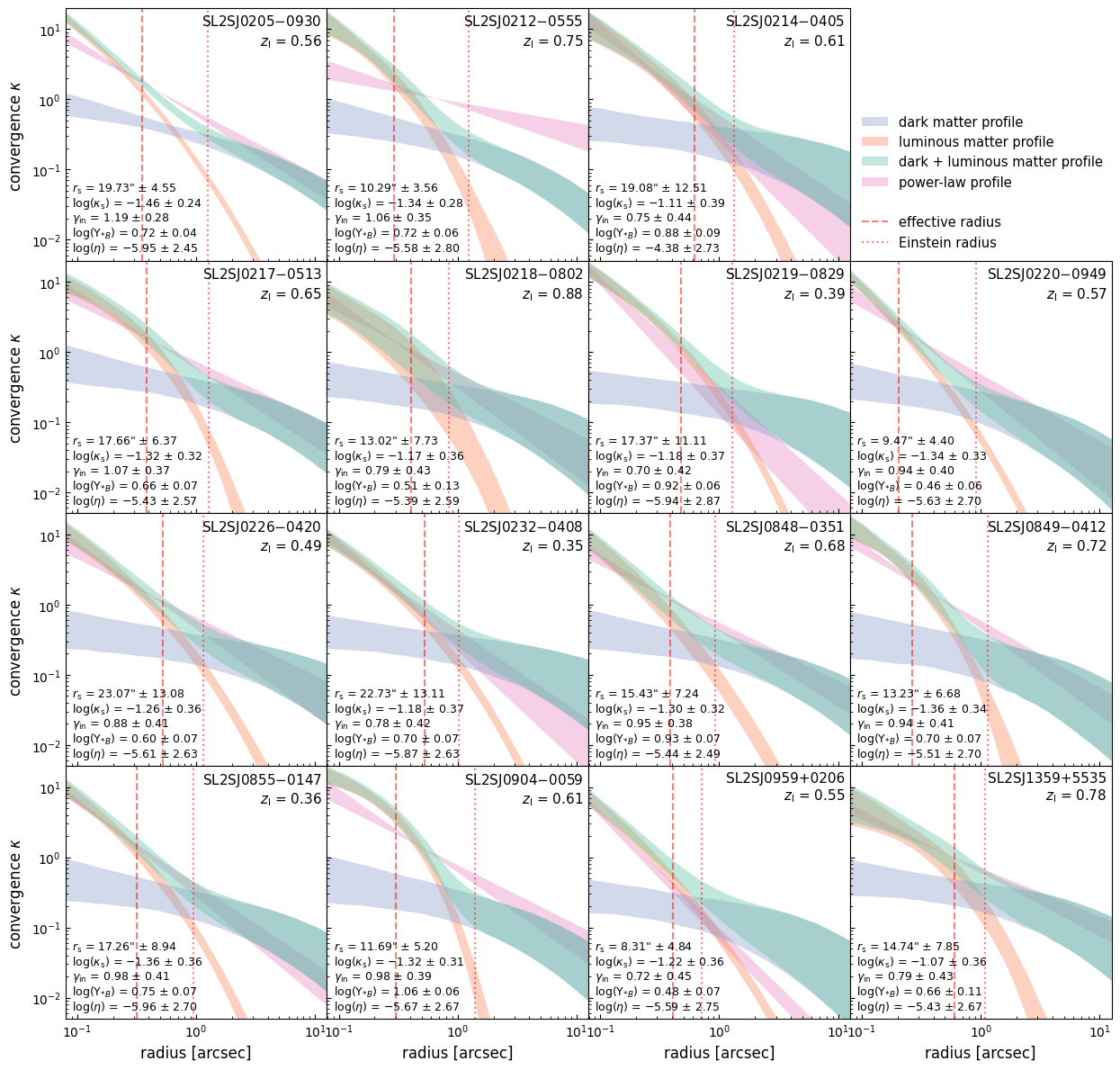}
 \caption{Luminous and dark matter decompositions for 15 (of 56) strong lensing systems used in our Bayesian hierarchical analysis.  The name and lens redshift are given at the top right, and the model parameter distributions in the bottom left for each subplot.  $R_s$ is the dark matter scale radius, $\kappa_s$ is the dark matter convergence at $R_s$, $\gamma_{\rm in}$ is the dark mater logarithmic inner slope, $\Upsilon_{*B}$ is the mass-to-light ratio in $B$-band in solar units, and $\eta$ is the mass-to-light gradient.  The remaining systems are shown in Figures~\ref{fig:decomp1}, \ref{fig:decomp2} and \ref{fig:decomp3}.}
 \label{fig:decomp0}
\end{figure*}

\begin{figure*}
 \includegraphics[width=1\linewidth]{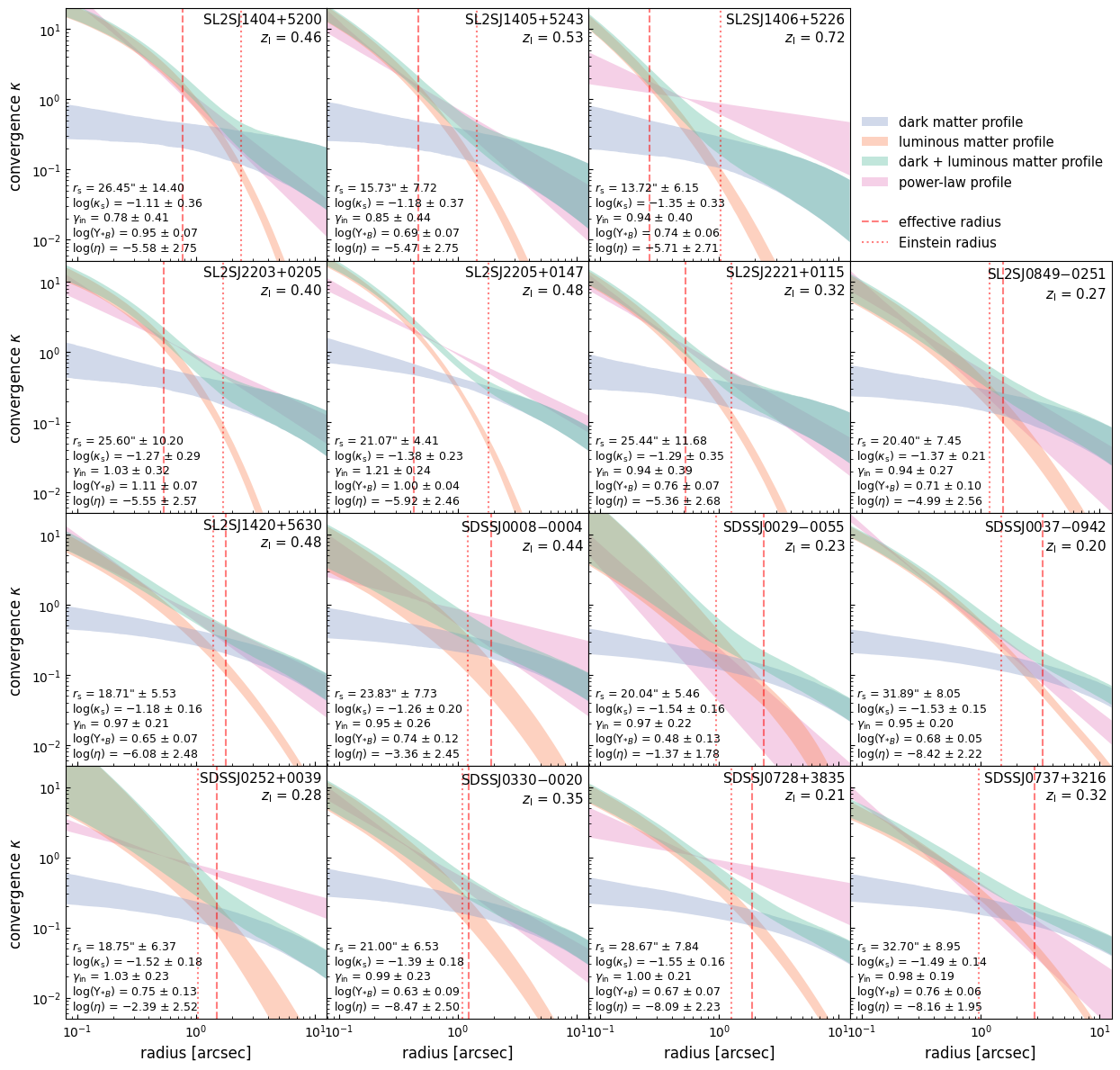}
 \caption{Luminous and dark matter decompositions for the next 15 (of 56) strong lensing systems used in our Bayesian hierarchical analysis.  See the caption of Figure~\ref{fig:decomp0} for the full description.}
 \label{fig:decomp1}
\end{figure*}

\begin{figure*}
 \includegraphics[width=1\linewidth]{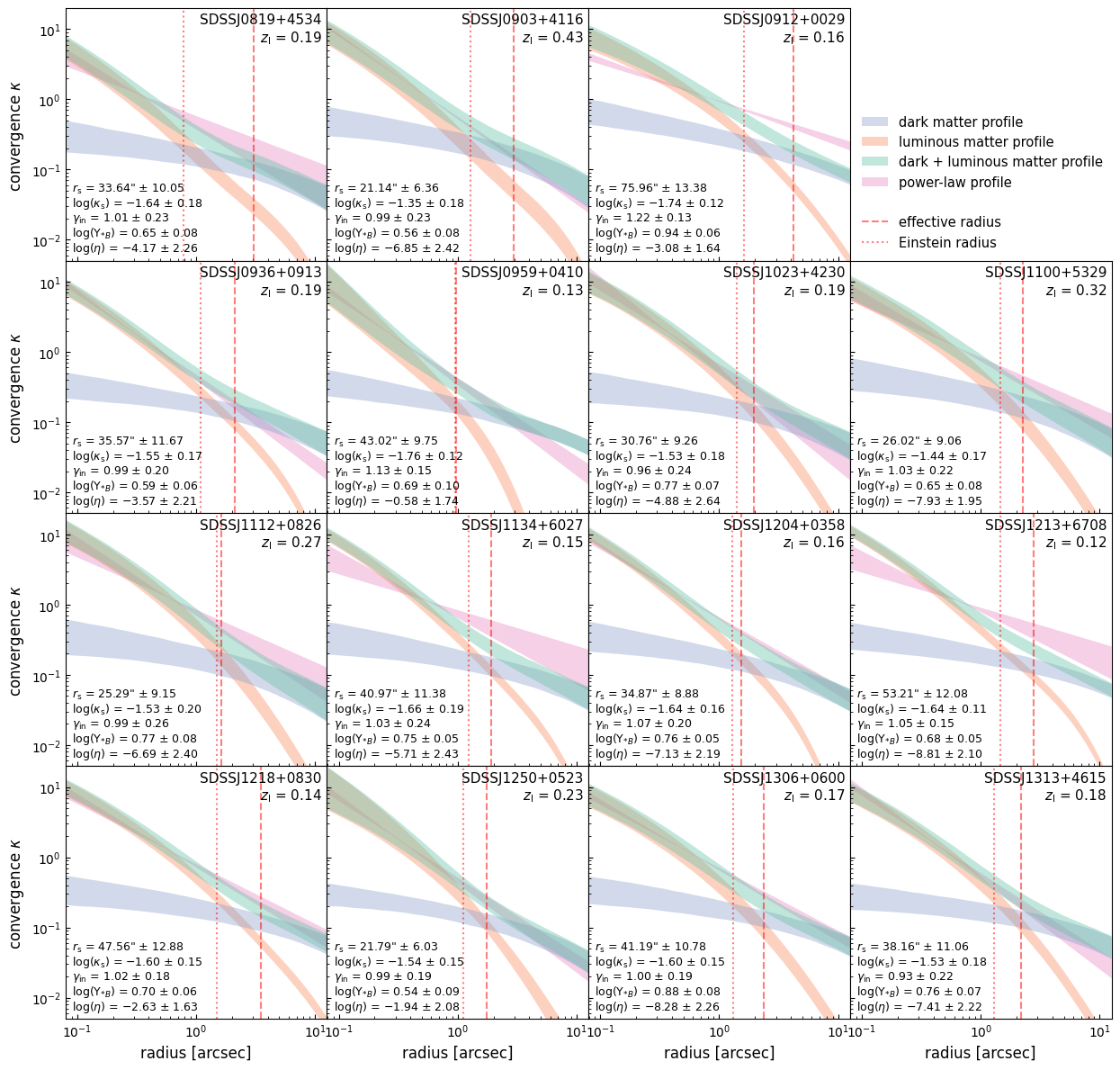}
 \caption{Luminous and dark matter decompositions for the next 15 (of 56) strong lensing systems used in our Bayesian hierarchical analysis.  See the caption of Figure~\ref{fig:decomp0} for the full description.}
 \label{fig:decomp2}
\end{figure*}

\begin{figure*}
 \includegraphics[width=1\linewidth]{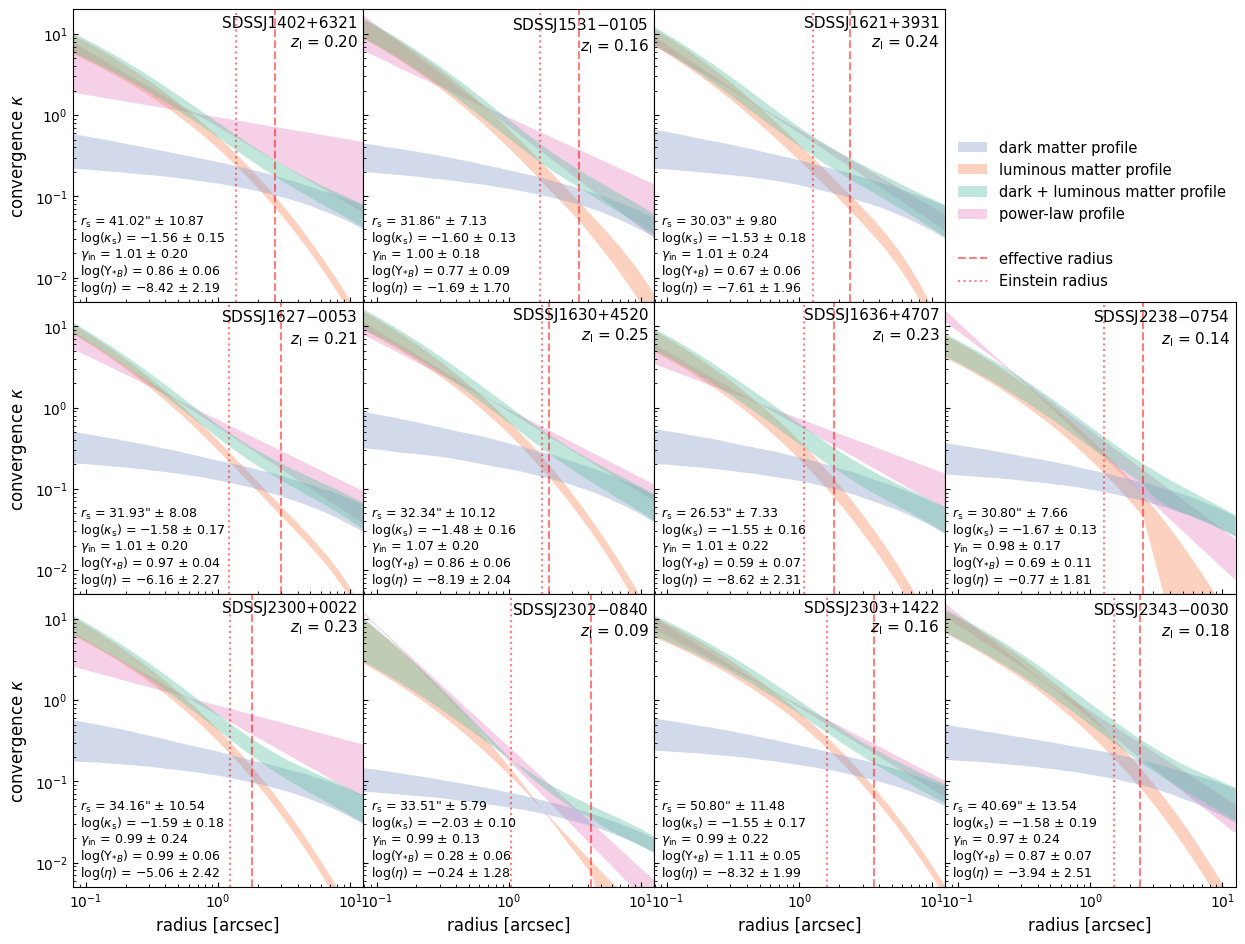}
 \caption{Luminous and dark matter decompositions for the last 11 (of 56) strong lensing systems used in our Bayesian hierarchical analysis.  See the caption of Figure~\ref{fig:decomp0} for the full description.}
 \label{fig:decomp3}
\end{figure*}

\section{Inner dark matter profile from IllustrisTNG}\label{appendix:illustris}

Here, we describe the extraction of the inner slope of the dark matter profile from halos in the IllustrisTNG simulation \citep{Nelson19}. We take the same 165 elliptical galaxy halos from \citet{Wang19}, which were selected to have the morphology of massive elliptical galaxies at both redshifts $z=0$ and $z=1$, and their $z=0$ stellar mass to satisfy $10^{10.7}\ \textrm{M}_{\odot} \leq M_{\star} \leq 10^{11.9}\ \textrm{M}_{\odot}$. This sample of elliptical galaxies represents passive evolution, where they have already quenched at $z > 1$.

We extract the dark matter particle distribution for each of these galaxies from the TNG100-1 run at snapshots $z =$ 0, 0.1, 0.2, 0.3, 0.4, 0.5, 0.7, and 1. We then extract the radial density profile of the dark matter by following the algorithm adopted by the \textsc{rockstar} catalog \citep[][see also \citealt{Anbajagane22}]{Behroozi13}. We bin the dark matter particles in each halo in equal-mass radial bins between $3\epsilon$ and $R_{200c}$, where $\epsilon = 0.74$ kpc is the gravitational softening length in the TNG100-1 run. We set the number of radial bins to either 50 or that needed for having at least 15 particles in each bin. We then take only the bins with \txr{$r < R_{200c}$}, where $R_{200c}$ is set by the $M_{200c}$ of each halo. %and the concentration $c$ is obtained from the mass--concentration relation from \citet{Ishiyama21}, as implemented in \textsc{colossus} \citep{Diemer18}. 
We then fit this inner radial density profile with a \txr{gNFW profile with free scale radius, amplitude, and inner logarithmic slope $\gamma_{\rm in}$.}

% with a power-law with free amplitude and exponent. We take the best-fit exponents as the inner logarithmic slope $\gamma_{\rm in}$ for each halo at the given snapshot or redshift. We do not fit a gNFW profile to the full radial profile extending up to $R_{200c}$ because, in that case, sometimes the concentration can end up extremely high or low or may result in bad fits for the whole profile if the concentration is fixed using the mass--concentration relation.

%\section{Hierarchical Analysis Velocity Dispersions Results} \label{appendix:vds}

%We present the observed and predicted velocity dispersion distributions for the 58 SL2S and SLACS strong lens sample (Figure~\ref{fig:vd}) used after applying our Bayesian hierarchical analysis, as discussed in \S\ref{sec:hierarchial}.

%%%%%%%%%%%%%%%%%%%%%%%%%%%%%%%%%%%%%%%%%%%%%%%%%%

% Don't change these lines
\bsp	% typesetting comment
\label{lastpage}
\end{document}